\documentclass[english]{article}
\usepackage[a4paper]{geometry}
\usepackage{babel}
\usepackage{verbatim}
\usepackage{mathtools}
\usepackage{booktabs}
\usepackage{enumitem}
\usepackage{bm}
\usepackage{amsmath}
\usepackage{amsthm}
\usepackage{amssymb}

\usepackage{color,xcolor}
\usepackage[numbers]{natbib}

\usepackage[figuresright]{rotating}
\usepackage{algorithm}
\usepackage{algpseudocode}
\usepackage{subcaption}
\usepackage{url}
\usepackage[pdfusetitle,
 bookmarks=true,bookmarksnumbered=false,bookmarksopen=false,
 breaklinks=false,pdfborder={0 0 1},backref=false,
 colorlinks=true, linkcolor=black, citecolor=black, urlcolor=black]{hyperref}
\usepackage{cleveref}
\usepackage{multirow}
\usepackage{setspace}

\makeatletter
\theoremstyle{definition}

\theoremstyle{plain}

\theoremstyle{remark}

\theoremstyle{definition}

\providecommand{\examplename}{Example}
\theoremstyle{plain}

\newtheorem{theorem}{Theorem}
\newtheorem{proposition}{Proposition}

\newcommand{\qbf}{\mathbf{q}}

\providecommand{\assumptionname}{Assumption}
\providecommand{\definitionname}{Definition}
\providecommand{\lemmaname}{Lemma}
\providecommand{\propositionname}{Proposition}
\providecommand{\remarkname}{Remark}
\providecommand{\theoremname}{Theorem}
\providecommand{\corollaryname}{Corollary}

\crefdefaultlabelformat{#2\textbf{#1}#3} %
\crefname{section}{\textbf{Section}}{\textbf{sections}}
\Crefname{section}{\textbf{Section}}{\textbf{Sections}}
\crefname{thm}{\textbf{Theorem}}{\textbf{theorems}}
\Crefname{thm}{\textbf{Theorem}}{\textbf{Theorems}}
\crefname{lem}{\textbf{Lemma}}{\textbf{lemmas}}
\Crefname{lem}{\textbf{Lemma}}{\textbf{Lemmas}}
\crefname{prop}{\textbf{Proposition}}{\textbf{Propositions}}
\Crefname{prop}{\textbf{Proposition}}{\textbf{Propositions}}
\crefname{algorithm}{\textbf{Algorithm}}{\textbf{Algorithms}}
\Crefname{algorithm}{\textbf{Algorithm}}{\textbf{Algorithms}}
\crefname{coro}{\textbf{Corollary}}{\textbf{corollaries}}
\Crefname{coro}{\textbf{Corollary}}{\textbf{corollaries}}
\crefname{defn}{\textbf{Definition}}{\textbf{definitions}}
\Crefname{defn}{\textbf{Definition}}{\textbf{definitions}}
\crefname{table}{\textbf{Table}}{\textbf{tables}}
\Crefname{table}{\textbf{Table}}{\textbf{tables}}
\crefname{figure}{\textbf{Figure}}{\textbf{figures}}
\Crefname{figure}{\textbf{Figure}}{\textbf{figures}}
\crefname{exple}{\textbf{Example}}{\textbf{examples}}
\Crefname{exple}{\textbf{Example}}{\textbf{examples}}
\Crefname{assumption}{\textbf{Assumption}}{\textbf{Assumptions}}
\crefname{assumption}{\textbf{Assumption}}{\textbf{Assumptions}}
\Crefname{rem}{\textbf{Remark}}{\textbf{Remarks}}
\crefname{rem}{\textbf{Remark}}{\textbf{Remarks}}

\newif\ifcomments
\commentsfalse 
\ifcomments
  \newcommand{\comm}[2][]{\textcolor{blue}{\textbf{[#1:} #2\textbf{]}}}
  \newcommand{\qd}[1]{\comm[QD]{#1}}
  \newcommand{\zl}[1]{\comm[ZL]{#1}}
\else
  \newcommand{\comm}[2][]{}
  \newcommand{\qd}[1]{}
  \newcommand{\zl}[1]{}
  \newcommand{\xx}[1]{}
\fi

\DeclareRobustCommand{\revise}[1]{#1}
\providecommand{\corollaryname}{Corollary}

\newcommand{\ee}{IPMO}

\makeatother

\begin{document}
\global\long\def\inprod#1#2{\left\langle #1,#2\right\rangle }%
\global\long\def\inner#1#2{\langle#1,#2\rangle}%
\global\long\def\binner#1#2{\big\langle#1,#2\big\rangle}%
\global\long\def\Binner#1#2{\Big\langle#1,#2\Big\rangle}%

\global\long\def\abs#1{|#1|}%

\global\long\def\norm#1{\lVert#1\rVert}%
\global\long\def\bnorm#1{\big\Vert#1\big\Vert}%
\global\long\def\Bnorm#1{\Big\Vert#1\Big\Vert}%

\global\long\def\setnorm#1{\Vert#1\Vert_{-}}%
\global\long\def\bsetnorm#1{\big\Vert#1\big\Vert_{-}}%
\global\long\def\Bsetnorm#1{\Big\Vert#1\Big\Vert_{-}}%

\global\long\def\brbra#1{\big(#1\big)}%
\global\long\def\Brbra#1{\Big(#1\Big)}%
\global\long\def\rbra#1{(#1)}%
\global\long\def\sbra#1{[#1]}%
\global\long\def\bsbra#1{\big[#1\big]}%
\global\long\def\Bsbra#1{\Big[#1\Big]}%
\global\long\def\cbra#1{\{#1\}}%
\global\long\def\bcbra#1{\big\{#1\big\}}%
\global\long\def\Bcbra#1{\Big\{#1\Big\}}%

\global\long\def\vertiii#1{\left\vert \kern-0.25ex  \left\vert \kern-0.25ex  \left\vert #1\right\vert \kern-0.25ex  \right\vert \kern-0.25ex  \right\vert }%

\global\long\def\matr#1{\bm{#1}}%

\global\long\def\til#1{\tilde{#1}}%
\global\long\def\wtil#1{\widetilde{#1}}%

\global\long\def\wh#1{\widehat{#1}}%

\global\long\def\mcal#1{\mathcal{#1}}%

\global\long\def\mbb#1{\mathbb{#1}}%

\global\long\def\mtt#1{\mathtt{#1}}%

\global\long\def\ttt#1{\texttt{#1}}%

\global\long\def\dtxt{\textrm{d}}%

\global\long\def\bignorm#1{\bigl\Vert#1\bigr\Vert}%
\global\long\def\Bignorm#1{\Bigl\Vert#1\Bigr\Vert}%

\global\long\def\rmn#1#2{\mathbb{R}^{#1\times#2}}%

\global\long\def\deri#1#2{\frac{d#1}{d#2}}%
\global\long\def\pderi#1#2{\frac{\partial#1}{\partial#2}}%

\global\long\def\limk{\lim_{k\rightarrow\infty}}%

\global\long\def\smid{\mskip1mu\mid\mskip1mu}%

\global\long\def\trans{\textrm{T}}%

\global\long\def\onebf{\mathbf{1}}%
\global\long\def\zerobf{\mathbf{0}}%
\global\long\def\zero{\mathbf{0}}%

\global\long\def\Euc{\mathrm{E}}%
\global\long\def\Expe{\mathbb{E}}%

\global\long\def\rank{\mathrm{rank}}%
\global\long\def\range{\mathrm{range}}%
\global\long\def\diam{\mathrm{diam}}%
\global\long\def\epi{\mathrm{epi} }%
\global\long\def\relint{\mathrm{relint} }%
\global\long\def\dom{\mathrm{dom}}%
\global\long\def\prox{\mathrm{prox}}%
\global\long\def\proj{\mathrm{Proj}}%
\global\long\def\for{\mathrm{for}}%
\global\long\def\diag{\mathrm{diag}}%
\global\long\def\and{\mathrm{and}}%
\global\long\def\var{\textrm{VaR}}%
\global\long\def\where{\mathrm{where}}%
\global\long\def\dist{\mathrm{dist}}%
\global\long\def\textop{\mathrm{op}}%
\global\long\def\level{\mathrm{lev}}%

\global\long\def\st{\mathrm{s.t.}}%

\global\long\def\inte{\operatornamewithlimits{int}}%
\global\long\def\cov{\mathrm{Cov}}%

\global\long\def\argmin{\operatornamewithlimits{arg\,min}}%
\global\long\def\argmax{\operatornamewithlimits{arg\,max}}%
\global\long\def\maximize{\operatornamewithlimits{maximize}}%
\global\long\def\minimize{\operatornamewithlimits{minimize}}%

\global\long\def\tr{\operatornamewithlimits{tr}}%

\global\long\def\dis{\operatornamewithlimits{dist}}%

\global\long\def\prob{{\rm Pr}}%

\global\long\def\spans{\textrm{span}}%
\global\long\def\st{\operatornamewithlimits{s.t.}}%
\global\long\def\subjectto{\textrm{subject to}}%

\global\long\def\Var{\operatornamewithlimits{Var}}%

\global\long\def\raw{\rightarrow}%
\global\long\def\law{\leftarrow}%
\global\long\def\Raw{\Rightarrow}%
\global\long\def\Law{\Leftarrow}%

\global\long\def\vep{\varepsilon}%

\global\long\def\dom{\operatornamewithlimits{dom}}%

\global\long\def\tsum{{\textstyle {\sum}}}%

\global\long\def\Cbb{\mathbb{C}}%
\global\long\def\Ebb{\mathbb{E}}%
\global\long\def\Fbb{\mathbb{F}}%
\global\long\def\Nbb{\mathbb{N}}%
\global\long\def\Rbb{\mathbb{R}}%
\global\long\def\Vbb{\mathbb{V}}%
\global\long\def\reals{\mathbb{R}}%

\global\long\def\extR{\widebar{\mathbb{R}}}%
\global\long\def\Pbb{\mathbb{P}}%

\global\long\def\Mrm{\mathrm{M}}%
\global\long\def\Acal{\mathcal{A}}%
\global\long\def\Bcal{\mathcal{B}}%
\global\long\def\Ccal{\mathcal{C}}%
\global\long\def\Dcal{\mathcal{D}}%
\global\long\def\Ecal{\mathcal{E}}%
\global\long\def\Fcal{\mathcal{F}}%
\global\long\def\Gcal{\mathcal{G}}%
\global\long\def\Hcal{\mathcal{H}}%
\global\long\def\Ical{\mathcal{I}}%
\global\long\def\Kcal{\mathcal{K}}%
\global\long\def\Lcal{\mathcal{L}}%
\global\long\def\Mcal{\mathcal{M}}%
\global\long\def\Ncal{\mathcal{N}}%
\global\long\def\Ocal{\mathcal{O}}%
\global\long\def\Pcal{\mathcal{P}}%
\global\long\def\Scal{\mathcal{S}}%
\global\long\def\Rcal{\mathcal{R}}%
\global\long\def\Tcal{\mathcal{T}}%
\global\long\def\Ucal{\mathcal{U}}%
\global\long\def\Wcal{\mathcal{W}}%
\global\long\def\Xcal{\mathcal{X}}%
\global\long\def\Ycal{\mathcal{Y}}%
\global\long\def\Zcal{\mathcal{Z}}%

\global\long\def\abf{\mathbf{a}}%
\global\long\def\bbf{\mathbf{b}}%
\global\long\def\cbf{\mathbf{c}}%
\global\long\def\dbf{\mathbf{d}}%
\global\long\def\ebf{\mathbf{e}}%
\global\long\def\fbf{\mathbf{f}}%
\global\long\def\gbf{\mathbf{g}}%
\global\long\def\pbf{\mathbf{p}}%
\global\long\def\qbf{\mathbf{q}}%
\global\long\def\sbf{\mathbf{s}}%
\global\long\def\lbf{\mathbf{l}}%
\global\long\def\ubf{\mathbf{u}}%
\global\long\def\vbf{\mathbf{v}}%
\global\long\def\wbf{\mathbf{w}}%
\global\long\def\xbf{\mathbf{x}}%
\global\long\def\ybf{\mathbf{y}}%
\global\long\def\zbf{\mathbf{z}}%
\global\long\def\Zbf{Z}%

\global\long\def\Abf{\mathbf{A}}%
\global\long\def\Ubf{\mathbf{U}}%
\global\long\def\Pbf{\mathbf{P}}%
\global\long\def\Ibf{\mathbf{I}}%
\global\long\def\Ebf{\mathbf{E}}%
\global\long\def\Mbf{\mathbf{M}}%
\global\long\def\Qbf{\mathbf{Q}}%
\global\long\def\Lbf{\mathbf{L}}%
\global\long\def\Pbf{\mathbf{P}}%
\global\long\def\Wbf{\mathbf{W}}%

\global\long\def\lambf{\bm{\lambda}}%
\global\long\def\mubf{\bm{\mu}}%
\global\long\def\alphabf{\bm{\alpha}}%
\global\long\def\sigmabf{\bm{\sigma}}%
\global\long\def\thetabf{\bm{\theta}}%
\global\long\def\deltabf{\bm{\delta}}%
\global\long\def\vepbf{\bm{\vep}}%
\global\long\def\pibf{\bm{\pi}}%
\global\long\def\phibf{\bm{\phi}}%

\global\long\def\abm{\bm{a}}%
\global\long\def\bbm{\bm{b}}%
\global\long\def\cbm{\bm{c}}%
\global\long\def\dbm{\bm{d}}%
\global\long\def\ebm{\bm{e}}%
\global\long\def\fbm{\bm{f}}%
\global\long\def\gbm{\bm{g}}%
\global\long\def\hbm{\bm{h}}%
\global\long\def\pbm{\bm{p}}%
\global\long\def\qbm{\bm{q}}%
\global\long\def\rbm{\bm{r}}%
\global\long\def\sbm{\bm{s}}%
\global\long\def\tbm{\bm{t}}%
\global\long\def\ubm{\bm{u}}%
\global\long\def\vbm{\bm{v}}%
\global\long\def\wbm{\bm{w}}%
\global\long\def\xbm{\bm{x}}%
\global\long\def\ybm{\bm{y}}%
\global\long\def\zbm{\bm{z}}%

\global\long\def\Abm{\bm{A}}%
\global\long\def\Bbm{\bm{B}}%
\global\long\def\Cbm{\bm{C}}%
\global\long\def\Dbm{\bm{D}}%
\global\long\def\Ebm{\bm{E}}%
\global\long\def\Fbm{\bm{F}}%
\global\long\def\Gbm{\bm{G}}%
\global\long\def\Hbm{\bm{H}}%
\global\long\def\Ibm{\bm{I}}%
\global\long\def\Jbm{\bm{J}}%
\global\long\def\Lbm{\bm{L}}%
\global\long\def\Obm{\bm{O}}%
\global\long\def\Pbm{\bm{P}}%
\global\long\def\Qbm{\bm{Q}}%
\global\long\def\Rbm{\bm{R}}%
\global\long\def\Sbm{\bm{S}}%
\global\long\def\Ubm{\bm{U}}%
\global\long\def\Vbm{\bm{V}}%
\global\long\def\Wbm{\bm{W}}%
\global\long\def\Xbm{\bm{X}}%
\global\long\def\Ybm{\bm{Y}}%
\global\long\def\Zbm{\bm{Z}}%
\global\long\def\lambm{\bm{\lambda}}%

\global\long\def\alphabm{\bm{\alpha}}%
\global\long\def\albm{\bm{\alpha}}%
\global\long\def\pibm{\bm{\pi}}%
\global\long\def\taubm{\bm{\tau}}%
\global\long\def\mubm{\bm{\mu}}%
\global\long\def\yrm{\mathrm{y}}%

\newcommand{\dd}{\mathop{}\!\mathrm{d}}
\newcommand{\tder}[2]{\frac{\dd #1}{\dd #2}}
\newcommand{\pder}[2]{\frac{\partial #1}{\partial #2}}

\title{Integrated Prediction and Multi-period Portfolio Optimization\footnote{YL and ZL contributed equally.}}

\author{%
Yuxuan Linghu\thanks{ Email: \texttt{ylinghu3@gatech.edu}, Georgia Institute of Technology.} \qquad
Zhiyuan Liu\thanks{Email: \texttt{zhiyuanliu@uchicago.edu}, The University of Chicago.} \qquad
Qi Deng\thanks{Email: \texttt{qdeng24@sjtu.edu.cn}, Shanghai Jiao Tong University.}
}

\date{November 17, 2026}

\maketitle

\begin{abstract}
Multi-period portfolio optimization is important for real portfolio management, as it accounts for transaction costs, path-dependent risks, and the intertemporal structure of trading decisions that single-period models cannot capture. However, multi-period portfolio optimization requires multi-stage return forecasts. Classical methods usually follow a two-stage framework: machine learning algorithms are employed to produce forecasts that closely fit the realized returns, and the predicted values are then used in a downstream portfolio optimization problem to determine the asset weights.
This separation leads to a fundamental misalignment between predictions and decision outcomes, while also ignoring the impact of transaction costs.
To bridge this gap, recent studies have proposed the idea of end-to-end learning, integrating the two stages into a single pipeline. 
This paper introduces \ee\ (Integrated Prediction and Multi-period Portfolio Optimization), a model for multi-period mean-variance portfolio optimization with turnover penalties. 
The predictor generates multi-period return forecasts that parameterize a differentiable convex optimization layer, \revise{and is trained using a mean-variance loss evaluated at the resulting turnover-regularized allocations.} For scalability, we introduce a mirror-descent fixed-point (MDFP) differentiation scheme that avoids factorizing the Karush--Kuhn--Tucker (KKT) systems \revise{and substantially reduces computational cost.}
In experiments with real market data and two representative time-series prediction models, \revise{\ee\ achieves higher full-period net returns and Sharpe ratios and lower turnover than the two-stage benchmarks under both tested predictors.}
Our results show that integrating machine learning prediction with optimization in the multi-period setting improves financial outcomes and remains computationally tractable.
\end{abstract}


\section{Introduction}
\label{section:intro}

The mean-variance optimization of \citet{markowitz1952} establishes the foundation of modern portfolio theory.
This framework has been widely applied to portfolio problems, where the optimal portfolio often maximizes a one-step-ahead objective.
Despite its simplicity, this greedy strategy may lead to suboptimal solutions in the long run.
A large out-of-sample study by \citet{demiguel2009optimal} demonstrates that iteratively applying single-period mean-variance optimization underperforms a simple equal-weight rule,
indicating the instability of this paradigm in real-world settings.

One extension of the Markowitz model is multi-period mean-variance optimization, which addresses multi-stage allocation decisions over a finite horizon.
Unlike single-period models, multi-period formulations incorporate transaction costs and path-dependent risk, making the problem both more realistic and significantly more challenging.
Classical work approaches this setting through stochastic programming. 
\citet{dantzig1993multi} formulate the multi-period portfolio problem as a multi-stage stochastic linear program and develop decomposition-based algorithms to handle the resulting scenario tree. However, these methods can be computationally intractable without distributional assumptions, as the branches grow exponentially with the horizon.
\citet{boyd2017} propose the model predictive control (MPC) approach for multi-period portfolio optimization, where the unknown parameters are replaced with their forecasted values.
At each time step, they obtain a multi-period allocation path with all available information to avoid unfavorable positions in the future, while only executing the first step.
The MPC approach divides the problem into two stages: parameter prediction and deterministic portfolio optimization, and therefore provides a tractable approximation to the original problem, where the two stages can be effectively handled separately.

In many fields, including portfolio optimization, it is well recognized that parameter estimation is crucial for downstream decision-making. 
Recent work shows that regularized estimators and shrinkage techniques can significantly reduce estimation risk and stabilize return and covariance forecasts, leading to more robust portfolios in high-dimensional settings \citep{kinn2018reducing, kircher2025optimal}.
Advancements in end-to-end, decision-focused learning provide novel frameworks to integrate these two stages in modeling. \citet{elmachtoub2022smart} develop the Smart ``Predict, then Optimize'' framework, which replaces standard statistical losses with a decision-aware objective that directly penalizes the suboptimality of the downstream linear program induced by prediction errors. They show that training with the decision-centric loss can yield models that exhibit lower predictive accuracy yet deliver better decision performance than those trained under conventional decoupled approaches.
However, solving the integrated optimization problems is computationally non-trivial in general. In multi-period settings, the number of decision variables in the optimization layer grows linearly with the planning horizon, posing challenges to the scalability of the integrated methods. Widely used differentiable optimization methods such as OptNet~\citep{optnet2017} and CvxpyLayer~\citep{agrawal2019differentiable} rely on differentiating through large KKT systems. As the planning horizon increases, the associated KKT systems grow rapidly in dimension, substantially inflating training time.

In this paper, we propose \ee, an integrated deep learning framework for multi-period portfolio optimization, and provide a tractable optimization algorithm for long-horizon decision-making. 
Building on the MPC formulation~\citep{boyd2017} and the decision-focused approach~\citep{elmachtoub2022smart}, we embed a multi-period mean-variance optimization layer directly into the learning model, connecting the predictions and allocation decisions. 
Instead of treating return prediction as an independent statistical task, \revise{the proposed framework trains return forecasts using a mean-variance loss through a turnover-regularized multi-period allocation layer.}
\revise{In empirical tests on real market data, \ee\ achieves higher full-period net returns and Sharpe ratios and lower turnover than two-stage training under both tested predictors, in sample and out of sample. These full-period gains coexist with regime-specific exceptions.}
In addition, to overcome computational limitations, we develop a mirror descent fixed-point (MDFP) differentiation algorithm that applies a single mirror descent step at the optimal solution to obtain a fixed-point equation. This algorithm circumvents the construction and inversion of large KKT matrices, substantially reducing training time in practice. Our contributions are as follows:
\begin{enumerate}
    \item We present \ee, a decision-focused rolling-horizon framework for multi-period mean-variance portfolio optimization with turnover costs. \revise{Empirical tests on real market data show higher full-period net returns and Sharpe ratios and lower turnover than two-stage training under both tested predictors.}
    \item We derive the MDFP algorithm, which obtains tractable gradients for a multi-stage optimization layer, enabling efficient differentiation without explicitly forming or inverting high-dimensional KKT systems. The resulting fixed-point formulation provides a general recipe for differentiating complex iterative solvers while maintaining scalability and low computational overhead across long planning horizons.
\end{enumerate}

The rest of the paper is organized as follows.
Section~\ref{section:review} reviews the previous work on portfolio optimization and decision-focused learning.
Section~\ref{section:background} explains the theoretical background of multi-period mean-variance optimization and introduces the conventional two-stage pipeline.
Section~\ref{section:dfl} develops the \ee\ framework, in which the predictor and the optimizer are jointly trained through the MDFP formulation and its efficient implicit differentiation scheme.
Section~\ref{section:exp} presents empirical results on real ETF data, \revise{comparing full-period performance and market-regime differences with the two-stage baseline and assessing computational scalability.}
Section~\ref{section:conclusion} concludes.

\section{Related work}
\label{section:review}

\paragraph{Portfolio optimization}
Portfolio optimization is one of the central problems in finance, formalized by \citet{markowitz1952} as the mean-variance optimization problem. The mean-variance model and its variants, such as the Black-Litterman model~\citep{black1992global} and the mean-semivariance model~\citep{estrada2007mean}, are popular among practitioners and academics. These frameworks maximize the risk-return trade-off under different metrics and constraints by selecting asset weights. In most empirical studies, the optimal portfolio is re-estimated periodically as the parameters change over time~\citep{jagannathan2003risk, brodie2009sparse, ledoit2017nonlinear}. Nonetheless, it is shown that the greedy single-period optimization may lead to suboptimal solutions over the long run~\citep{demiguel2009optimal}. \citet{garleanu2013} show that in the presence of transaction costs, the optimal strategy is to gradually rebalance towards a weighted average of the current target portfolio and the expected future target portfolios, highlighting the necessity of multi-stage planning. 

There are various methods to approach multi-period portfolio optimization. Building on the classical stochastic programming formulation~\citep{dantzig1993multi}, subsequent work optimizes expected terminal wealth or the utility of terminal performance over the investment horizon~\citep{zhou2000continuous}. 
However, such objectives lead to time-inconsistent decisions, as the allocation may not remain optimal as time progresses~\citep{li2000optimal}. Instead of applying a static pre-commitment strategy, later studies further examine multi-period mean-variance optimization in a dynamic setting~\citep{basak2010dynamic}. \citet{yu2020neural} fit a neural network to produce multi-period returns from simulation of economic factors. The return predictions are passed into a Mean-CVaR optimizer to obtain portfolio weights. Building on the model predictive control (MPC) formulation~\citep{boyd2017}, several studies adopt related ``predict-then-optimize'' schemes for multi-period allocation~\citep{oprisor2020multi, li2022multi}, where forecasts of returns and risks are passed into a deterministic optimization layer at each rebalancing date. Related sequential models combine state-dependent return estimates with online portfolio selection and explicit transaction costs, while retaining a modular estimator-optimizer pipeline~\citep{guo2023online}. On the other hand, some studies rely on sophisticated methods such as deep reinforcement learning without an explicit optimization process to generate portfolio weights directly~\citep{cui2024multi, jiang2025high}. Motivated by the limitations of both pipelines, a complementary direction makes the optimizer part of the learning model, so that forecasts are shaped directly by the portfolio objective.

\paragraph{Decision-focused learning}
The decision-focused line in portfolio optimization builds on differentiable optimization. \citet{optnet2017} introduce OptNet to embed quadratic programs as neural layers via implicit differentiation of the KKT system, and \citet{agrawal2019differentiable} generalize this capability to disciplined convex programs through CvxpyLayer. \citet{blondel2022efficient} present a general method of applying automatic implicit differentiation for a broad range of optimization problems. More recent studies further improve the computational tractability of large-scale end-to-end learning. BPQP~\citep{pan2024bpqp} simplifies the backward pass as a QP and adopts efficient solvers for the forward and the backward passes separately. This decoupling provides greater flexibility in selecting solvers and accelerates the optimization process. dQP~\citep{magoon2025dqp} computes the active constraint set to prune the KKT system, reducing computation costs for large-scale sparse problems. Alongside differentiable layers, recent portfolio studies align forecasts with investment decisions through alternative mechanisms. \citet{sarkar2025integration} train an optimization-aware regression tree under a mean-variance objective to reduce downstream allocation error, while \citet{xidonas2025integration} combine forward-looking forecasts with multiple-criteria rankings for portfolio construction. Building on these methodological streams, finance studies move the portfolio optimization objective inside the learning loop in diverse settings. \citet{lee2024anatomy} study how decision-focused learning reshapes return predictors to improve decision quality, and complementary end-to-end GMV results are reported by \citet{bongiorno2025end}. \citet{Uysal2024} optimize risk-budgeting objectives in an end-to-end manner, demonstrating significant out-of-sample performance gains from including the optimization layer over the model-free approach. Tail- and risk-sensitive variants have primarily been explored through distributionally robust, single-period end-to-end formulations \citep{costa2023distributionally}. For tradability, single-period cardinality-constrained portfolios are trained in an integrated, predict-and-optimize fashion~\citep{ANIS2025739}. Collectively, these works establish single-period end-to-end portfolios as a rigorous testbed, and provide the building blocks for multi-period end-to-end formulations.

Recent studies have begun to extend this literature to multi-period settings. \citet{zhang2025end} develop a two-branch policy network that directly produces portfolio weights and asset-specific stop-loss levels over a multi-period investment process. In a different setting, \citet{bae2026decision} apply decision-focused learning to multi-stage goal-based investing through surrogate learning for a robust goal-programming model. These studies demonstrate the broader scope of multi-period end-to-end learning, but respectively focus on direct policy learning and goal-based planning rather than training return forecasts through a receding-horizon mean-variance layer with turnover-coupled decisions.

It is recognized that scaling portfolio optimization models to realistic asset universes and long decision horizons raises computational challenges due to the increasing dimensionality of the resulting optimization problems. In the cross-section, \citet{Bertsimas2022} propose a sparse portfolio selection method that delivers substantial speedups at real-world scales. In end-to-end optimization, it is well recognized that embedding a differentiable optimization layer into the portfolio model incurs additional computational cost~\citep{ANIS2025739}, posing challenges to the scalability of the end-to-end optimization. Multi-period formulations further amplify the computational burden, with costs growing at least linearly in the horizon. \citet{WAHLBERG201283} apply ADMM for multi-period portfolio optimization problems with transaction costs and yield orders-of-magnitude speedups over generic solvers.

The above literature highlights both the benefits of multi-period portfolio optimization and the potential of decision-focused learning for improving portfolio decisions. 
However, to the best of our knowledge, existing work has not developed a decision-focused rolling-horizon mean-variance framework that trains return forecasts through an explicit convex multi-period optimization layer and differentiates the realized portfolio loss through a turnover-coupled allocation path.
Furthermore, current differentiable optimization approaches predominantly rely on implicit differentiation through large KKT systems. Consequently, the computational costs of the backward pass can become the main bottleneck of the integrated methods as the system size grows.
These limitations highlight the need for a computationally tractable mechanism that enables multi-period, decision-driven learning without incurring the full cost of differentiating a large-scale constrained optimization problem.

\section{Background}

\label{section:background}
Consider a long-only portfolio of $N$ traded assets at discrete decision dates $t{+}1,\ldots,t{+}H$.
At the start date $t$, the portfolio holds a known weight vector $z_t\in\mathbb{R}^N$.
We plan a path of future portfolio allocations
$\Zbf_t=(z_{t+1},\ldots,z_{t+H})$ over an $H$-day horizon.
We denote $Z_{t,i}=z_{t+i}$ as the $i$-th stage allocation, $i=1,\ldots,H$.
We start from a general optimization problem that maximizes intertemporal utility, subject to feasibility
\[
\label{eq:general-utility}
\max_{\Zbf_t}
\ \mathbb{E}\left[\mathcal{U}(\Zbf_t) \,\middle|\, \mathcal{F}_t\right],
\quad \text{s.t.}\quad \Zbf_t\in\boldsymbol{\Omega},
\]
where $\mathcal{U}(\Zbf_t)$ denotes the utility function that aggregates expected portfolio returns, risks, and transaction costs along the $H$-period investment horizon, $\mathcal{F}_t$ denotes the available information up to day $t$, and $\boldsymbol{\Omega}$ denotes the feasible set of portfolio allocations. Typically, feasibility on each date is given by a simplex of the long-only portfolio weights
$\Omega_s = \{\, z \in \mathbb{R}^N \mid \mathbf{1}^\top z = 1,\, z \ge 0 \,\}$,
and the multi-period feasible set is
$\boldsymbol{\Omega} = \Omega_{t+1} \times \cdots \times \Omega_{t+H}$. To obtain a tractable objective, we assume a time-separable structure~\citep{lezmi2022multi}:
\[
\label{eq:separable}
-\,\mathbb{E}\left[\mathcal{U}(\Zbf_t) \,\middle|\, \mathcal{F}_t\right]
= \sum_{s=t+1}^{t+H}
\big[
g_s(z_s) + \lambda h_s(z_s-z_{s-1})\big],
\]
where $g_s:\mathbb{R}^N\to\mathbb{R}$ and $h_s:\mathbb{R}^N\to\mathbb{R}$ may depend on $\mathcal{F}_t$. The objective is additive across the planning horizon. At each stage, the expected objective decomposes into a static term determined by the current portfolio and a dynamic term depending on the rebalancing from the previous portfolio. 
Trading frictions can be internalized through a turnover-based regularizer applied to the rebalancing vector $z_s-z_{s-1}$ scaled by $\lambda\ge 0$. 
This formulation captures the intertemporal trade-offs that matter in execution while keeping the constraint set minimal and realistic.

A commonly adopted specification for $g_s$ is the MPC mean-variance objective.
Let $y_{s}\in\mathbb{R}^N$ denote the future return vector
and $V_{s}\in\mathbb{S}^N_{++}$ denote the positive definite covariance matrix, where $s\in\{t{+}1,\ldots,t{+}H\}$. Let $\hat{y}_s$ and $\hat{V}_s$ denote their forecasted values, respectively.
The mean-variance objective for a single period is
\begin{equation}
\label{eq:mv-single}
g_s(z_s,\mathcal{F}_t) = \frac{\delta}{2}\,z_s^\top \hat{V}_{s} z_s - \hat{y}_{s}^\top z_s,
\end{equation}
where $\delta$ is the risk parameter. This formulation preserves convexity because $\hat{V}_{s}\succ 0$, and it rewards exposure to higher expected returns while penalizing forecasted variance. 

To mitigate excessive portfolio turnover, we introduce a penalty on trading activity. While the standard $L_1$ penalty $\|z_s-z_{s-1}\|_1$ is widely employed~\citep{hautsch2019large}, its non-differentiability at the origin poses challenges for first- and second-order numerical optimization methods. To ensure a smooth and well-conditioned objective, we substitute the absolute value term $|\cdot|$ with a differentiable approximation
\[
\rho_\kappa(x)=\sqrt{x^2+\kappa},
\]
where $\kappa>0$ is a small perturbation parameter that controls the degree of smoothing.
The function $\rho_\kappa$ is twice differentiable, with its gradient given by $\nabla\rho_\kappa(x)={x}/{\sqrt{x^2+\kappa}}$.
These properties yield stable gradients and bounded curvature, facilitating robust numerical optimization. For convenience, we define
$\rho(v)=\sum_{i=1}^N \rho_\kappa(v_i)$ where $v\in\mathbb{R}^N$.
Incorporating this smooth penalty, the multi-period mean-variance problem is formulated as
\begin{equation}
\label{eq:mv-multip}
\begin{aligned}
\min_{\Zbf_t}
& \sum_{s=t+1}^{t+H}
\left[
\frac{\delta}{2}\,z_s^\top \hat{V}_{s}z_s
-\hat{y}_{s}^\top z_s
+\lambda\,\rho(z_s-z_{s-1})
\right],
\\[2pt]
\text{s.t.} & \quad
\Zbf_t\in\boldsymbol{\Omega}.
\end{aligned}
\end{equation}

In the conventional two-stage framework, the forecasting and decision-making processes are conducted independently. This procedure is detailed in \cref{alg:pto}.
In the forecasting stage, a model $\phi_\theta$, parameterized by $\theta\in\Theta$, is trained to predict future returns, and the downstream allocation problem is then solved based on the predicted values. 
Let $X_s \in \mathbb{R}^{L\times N}$ denote the past $L$-day returns for all $N$ assets on day $s$. Let $Y_s = (y_{s+1},\ldots,y_{s+H}) \in \mathbb{R}^{H\times N}$ denote the realized returns from day $s+1$ to $s+H$, and $\hat{Y}_s = \phi_\theta(X_s)$ the predicted returns.
On day $t$, the model is trained using a rolling look-back window of realized returns $\{(X_s, Y_s), s=t-T-H+1, \dots, t-H\}$ with a prediction loss $\ell_{\mathrm{p}}\big(\hat{Y}_s,\, Y_s\big)$ plus regularization terms.
In practice, the prediction loss $\ell_{\mathrm{p}}$ is often the mean squared error and is estimated on mini-batches. 
This leads to the following training problem:
\begin{equation}
\label{eq:forecast-training}
\min_{\theta \in \Theta} \;  \mathcal{L}_{\mathrm{p}}(\theta)
=
\frac{1}{T}
\sum_{s = t-T-H+1}^{t-H}
\ell_{\mathrm{p}}(\hat{Y}_s,\, Y_s)
+
\beta R(\theta),
\end{equation}
where $R(\theta)$ denotes a regularization term weighted by the coefficient $\beta \geq 0$.
The optimization problem~\eqref{eq:forecast-training} is typically solved using a learning algorithm such as Adam~\citep{kingma2015adam}, which yields the optimized parameters $\theta^*$.
On a new decision date $t$, the trained model $\phi_{\theta^*}$ generates the predicted returns   $\hat{Y}_{t}$ for the next $H$ periods. The predicted per-period covariance matrix $\hat{V}_{s}, s=t+1,\ldots,t+H$ is estimated based on a rolling window of past returns.
To ensure the covariance matrix remains positive definite, we add a small perturbation term $\varepsilon I$ to $\hat{V}_{s}$. 
An optimizer is then applied to solve \eqref{eq:mv-multip} using these inputs to obtain a path of allocations $\Zbf_t^*$. In the MPC approach, the investor executes the first step $z^*_{t+1}$.

In \cref{alg:pto}, the training loss for the prediction model is agnostic to the structure and objectives of the downstream optimizer. Though computationally tractable, the separation of the forecasting and decision-making stages can lead to suboptimal portfolio choices. For instance, small forecast errors can lead to materially different portfolio choices, as demonstrated by the well-documented flaw in the Markowitz model~\citep{best1991sensitivity}. In the multi-period setting, this problem is exacerbated by the difficulty of multi-period forecasting and the path dependence of the downstream optimization problem. Moreover, transaction costs are not considered in the forecasting stage. Frequent fluctuations in predictions can significantly diminish portfolio performance, especially when the benefits of rebalancing are largely offset by transaction costs.

\begin{algorithm}[t]
\caption{Two-stage forecasting and portfolio optimization}
\label{alg:pto}
\begin{algorithmic}
\Require Training data $\{(X_s, Y_s), s=t-T-H+1,\dots,t-H\}$, risk parameter $\delta$, penalty $\lambda$, smoothing factor $\kappa$, regularizer $\beta$;
\Ensure First-step allocation $z^*_{t+1}$;
\State \hspace{-0.9em}\textbf{Stage I (Forecasting)}
\State {Run a training algorithm, such as Adam, to solve \eqref{eq:forecast-training} and obtain a solution $\theta^*$};
\State \hspace{-0.9em}\textbf{Stage II (Decision-making)}
\State Compute $\hat{Y}_t=\phi_{\theta^*}(X_t)$;
\State Compute $\hat{V}_s,\ s=t+1,\ldots,t+H$ based on past returns;
\State Solve \eqref{eq:mv-multip} to obtain  $\Zbf_t^*$ and execute the first-step portfolio $z^*_{t+1}$.
\end{algorithmic}
\end{algorithm}

\section{\ee\ learning framework}

\label{section:dfl}

\subsection{Problem setup}

In this section, we introduce the Integrated Prediction and Multi-period Portfolio Optimization (\ee) framework, which updates the prediction model based on the quality of the induced decisions. 
Given the input $X_s$, the model first generates an intermediate variable $\hat{Y}_s=\phi_\theta(X_s)\in\mathbb{R}^{H\times N}$. This variable defines the multi-period optimization objective in \eqref{eq:mv-multip}, denoted by $\hat{F}_s$, where $\hat{Y}_s$ serves as the predicted returns. The following optimization layer solves this problem to obtain the optimal multi-period allocation $\Zbf_s^*(\theta)$. The output variable $\Zbf_s^*(\theta)$ is evaluated against the realized returns $Y_s$ by a decision loss $\ell_{\mathrm{d}}(\Zbf_s^*(\theta),Y_s)$, which measures the performance of the resulting portfolio.
The general integrated learning objective is
\begin{equation}
\label{eq:tempo-abstract}
\begin{aligned}
\min_{\theta \in \Theta}
&\;
\frac{1}{T}\sum_{s = t-T-H+1}^{t-H}
\ell_{\mathrm{d}}\bigl(\Zbf_s^*(\theta),\,Y_s\bigr)
, \\[2pt]
\text{s.t.} &\;
\Zbf_s^*(\theta)
=
\arg\min_{\Zbf_s\in\boldsymbol{\Omega}}\,
\hat{F}_s \bigl(\Zbf_s,\,\hat{Y}_s\bigr).
\end{aligned}
\end{equation}
Unlike the two-stage model, \eqref{eq:tempo-abstract} does not evaluate the predicted returns $\hat{Y}_s$ against the true returns $Y_s$.
Intuitively, the model parameters $\theta$ are chosen to minimize the expected decision loss induced by the downstream optimizer. Since the model $\phi_\theta$ only affects the objective through the solution $\Zbf_s^*(\theta)$, the prediction accuracy is rewarded only to the extent that it improves the portfolio performance.

The integrated multi-period mean-variance optimization problem takes the bilevel form
\begin{equation}
\label{eq:tempo-mv-final}
\begin{aligned}
\min_{\theta}\quad
& \frac{1}{T}\sum_{s=t-T-H+1}^{t-H}
  \frac{1}{H}\sum_{k=s+1}^{s+H}
  \left[
    -z_k^{*}(\theta)^\top y_k
    +\frac{\delta}{2}\,z_k^{*}(\theta)^\top V_k z_k^{*}(\theta)
  \right],
\\[2pt]
\text{s.t.}\quad
& \Zbf_s^{*}(\theta)
  =\arg\min_{\Zbf_s\in\boldsymbol{\Omega}}
  \sum_{k=s+1}^{s+H}
  \left[
    \frac{\delta}{2}\,z_k^\top \hat{V}_{k}z_k
    -\hat{y}_{k}^\top z_k
    +\lambda\,\rho(z_k-z_{k-1})
  \right],
  \qquad \text{where } z_s=Z_{s-1,1}^*.
\end{aligned}
\end{equation}
The covariance input $\hat{V}_k$ is a sample covariance estimate constructed from historical return observations available at the decision date.
\revise{The outer problem trains the predictor using the horizon-wise mean-variance loss of the induced allocations evaluated with realized returns. The inner problem balances predicted returns and variance with a turnover penalty, allowing transaction costs to influence training through the allocation layer.}

\subsection{Methodology}

The previous subsection presents the \ee\ objective with an inner multi-stage portfolio program and an outer \revise{mean-variance training loss}.
Training such a model requires propagating gradients from the evaluation loss back to the predictor parameters.
The key difficulty lies in differentiating through the inner optimizer: the gradient of the outer loss with respect to $\theta$ depends on the Jacobian of the solution $\Zbf_t^{*}(\theta)$.
For convex programs with simplex constraints and turnover penalties, computing this Jacobian analytically is intractable.
Existing differentiable optimization frameworks, such as OptNet~\citep{optnet2017}, CvxpyLayer~\citep{agrawal2019differentiable}, and BPQP~\citep{pan2024bpqp}, tackle this by applying the implicit-function theorem to the KKT system.
While theoretically exact, this approach requires solving large linear systems involving KKT matrices at every backward step, which incurs substantial computational and memory costs.
Each gradient evaluation entails matrix factorizations whose complexity scales cubically with problem size, and storing intermediate solver states quickly becomes prohibitive in multi-stage or high-dimensional settings.
As a result, KKT-based differentiation often dominates runtime and limits scalability in \ee\ training.

In contrast, we do not differentiate through the KKT system directly.
Instead, we reinterpret the optimality conditions of the inner convex problem as a fixed-point equation induced by a mirror-descent map on the simplex.
Concretely, the entropic mirror map yields a softmax-type multiplicative update at each stage, whose normalization operator implicitly enforces both the equality and nonnegativity constraints.  
The normalization is blockwise across stages, while the gradient entering each block retains the local temporal dependence induced by the turnover penalty.
As a result, the KKT multipliers never need to be formed explicitly and the sensitivity system can be evaluated through sparse, local-in-time Jacobian-vector products.
This fixed-point formulation avoids both explicit argmin differentiation and the unrolling of first-order solvers: the equilibrium allocation $\Zbf_t^*(\theta)$ is characterized as the fixed point of the MD map, and differentiating through this relation yields efficient implicit gradients without constructing or factorizing the KKT matrix.  
Our approach therefore combines the geometric interpretability of mirror descent with the computational efficiency of fixed-point implicit differentiation, providing a scalable alternative to traditional KKT-based layers.

\subsubsection{Mirror descent}
Mirror Descent (MD)~\citep{nemirovskij1983problem} is a first-order algorithm for constrained convex optimization that adapts its update step to the underlying geometry of the feasible set. It generalizes classical gradient descent by employing non-Euclidean distance measures, allowing optimization to respect the structural properties of the constraint set.

Let \(f:\boldsymbol{\Omega}\to\mathbb{R}\) be a differentiable convex function, and let \(\mathcal{C}\subseteq\mathbb{R}^N\) denote the closed convex feasible set at each stage.
Given the $m$-th iterate $\Zbf_t^m\in\boldsymbol{\Omega}$, let $g_s^m:=\nabla_{z_s}f(\Zbf_t^m)$ denote the block gradient with respect to the $s$-th-stage allocation, where $s=t+1,\ldots,t+H$.
MD defines the next point as the minimizer of a local first-order model of $f$ regularized by a Bregman divergence:
\begin{equation}
\label{eq:md-primal-general}
z_s^{m+1}
=
\arg\min_{w\in\mathcal{C}}
\Big\{
\langle g_s^{m},\,w\rangle
+\frac{1}{\eta}\,D_\psi(w,z_s^m)
\Big\},
\end{equation}
where $w\in\mathcal{C}$ is the optimization variable, $\eta>0$ is the step size and $\psi:\mathcal{C}\to\mathbb{R}$ is a strictly convex mirror map that induces the Bregman divergence $D_\psi(u,v)=\psi(u)-\psi(v)-\langle\nabla\psi(v),u-v\rangle$, where $u,v\in\mathcal{C}$ are arbitrary points in the domain.
The MD step \eqref{eq:md-primal-general} thus computes a proximal minimizer of a linearized objective under a geometry defined by $\psi$.
For optimization over the probability simplex, a natural mirror map is the negative entropy $\psi(z_s)=\sum_{i=1}^N z_{s,i}\log z_{s,i}$,
which induces the Kullback-Leibler (KL) divergence   
$D_\psi(u,v)
=\sum_{i=1}^N u_i\log\frac{u_i}{v_i}
-\sum_{i=1}^N(u_i-v_i)$. Here, $i$ denotes the $i$-th asset.
With this choice, the MD update admits a closed form:
\begin{equation}
\label{eq:md-simplex-update}
z^{m+1}_{s,i}
=\frac{z^m_{s,i}\exp(-\eta g_{s,i}^m)}
{\sum_{j=1}^N z^m_{s,j}\exp(-\eta g_{s,j}^m)},
\qquad i=1,\ldots,N.
\end{equation}
This step can be interpreted as performing a single mirror-proximal solve of a convex subproblem, providing a differentiable mapping that approximates the exact $\arg\min$ operator.
Because feasibility is preserved automatically, the update is well suited for our setting. 
For a single-stage problem on the simplex, it is immediate that any optimal point is a fixed point of the MD map. 
The following theorem extends this property to the multi-stage case, where the feasible set is a product of simplices. The proof is provided in~\ref{app:proof-fixedpoint}.

\begin{theorem}\label{thm:md-fixedpoint}
Let $f:\boldsymbol{\Omega}\to\mathbb{R}$ be differentiable and strongly convex.  
Then the mirror descent update applied stagewise as in~\eqref{eq:md-simplex-update} satisfies the condition that the optimal solution $\Zbf_t^{*}$ remains invariant under one iteration.
\end{theorem}

The fixed-point characterization therefore holds for the entire multi-stage allocation problem, 
providing a convenient basis for implicit differentiation through the inner solver.

\subsubsection{Implicit differentiation of the MD fixed point}

Building on~\cref{thm:md-fixedpoint}, we now use the MD fixed point  to construct a differentiable surrogate for the inner optimization. 
The MD update at each stage takes the form
$\Phi_s(\Zbf_t,\theta)
=\operatorname{Normalize}\Big(
 z_s \odot \exp\big(-\eta\,\nabla_{z_s}\hat{F}_t(\Zbf_t,\hat{Y}_t)\big)
\Big), s=t+1,\ldots,t+H$, where $\nabla_{z_s}\hat{F}_t(\Zbf_t,\hat{Y}_t)$
denotes the gradient with respect to the $s$-th stage allocation, and $\odot$ denotes the Hadamard (elementwise) product.
We collect $\Phi(\Zbf_t,\theta)=(\Phi_{t+1},\ldots,\Phi_{t+H})$.
By~\cref{thm:md-fixedpoint}, the optimal matrix $\Zbf_t^{*}$ satisfies the fixed-point relation
\begin{equation}
\label{eq:md-fp}
\Zbf_t^{*}=\Phi(\Zbf_t^{*},\theta),
\qquad \Zbf_t^{*}\in\boldsymbol{\Omega},
\end{equation}
which serves as a differentiable surrogate for the exact $\arg\min$. Differentiating this fixed-point relation with respect to $\theta$ yields the implicit gradient expression in~\cref{prop:md-gradient}, and the full derivation is provided in~\ref{app:proof-md-gradient}.
\begin{proposition}
\label{prop:md-gradient}
Let $\partial_{\Zbf}\Phi(\Zbf,\theta) \in\mathbb{R}^{(NH)\times(NH)}$ 
denote the Jacobian of $\Phi$ with respect to $\Zbf$ and suppose $I - \partial_{\Zbf_t^{*}}\Phi(\Zbf_t^{*},\theta)$ is invertible.
Then, we have
\begin{equation}
\label{eq:md-gradient}
\frac{\partial \Zbf_t^{*}}{\partial \theta}
=\big(I - \partial_{\Zbf_t^{*}} \Phi(\Zbf_t^{*},\theta)\big)^{-1}
\,\frac{\partial \Phi(\Zbf_t^{*},\theta)}{\partial \theta}.
\end{equation}
\end{proposition}

  We note that the Jacobian in~\eqref{eq:md-gradient} is a partial derivative and therefore does not account for the path along which $\theta$ influences $\Zbf_t^{*}$ through the preceding allocation $\Zbf_{t-1}^{*}$.
  The complete gradient follows from $\frac{\dd \Zbf_t^{*}}{\dd \theta} = \frac{\partial \Zbf_t^{*}}{\partial \theta} + \frac{\partial \Zbf_t^{*}}{\partial \Zbf_{t-1}^{*}}\frac{\dd \Zbf_{t-1}^{*}}{\dd \theta}$, in which the two terms represent the direct and the indirect dependence of $\Zbf_t^{*}$ on $\theta$, respectively.
  Since evaluating the indirect term would require differentiating through the entire coupled sequence of rolling optimization problems, at a computational cost that grows with the length of the training window, we truncate the recursion during backpropagation and retain only the direct term $\frac{\partial \Zbf_t^{*}}{\partial \theta}$ obtained in~\cref{prop:md-gradient}.

The expression in~\eqref{eq:md-gradient} demonstrates that the gradient of the fixed-point solution can be computed without explicitly differentiating through the KKT system. 
Although boundary solutions are covered under strict complementarity in~\cref{thm:fp-kkt}, near-degenerate active sets and coordinates at numerical precision can still make fixed-point sensitivities ill-conditioned~\citep{optnet2017}. 
To mitigate these finite-precision effects, we impose a small lower bound of $10^{-8}$ on the coordinates in the numerical implementation.
We next show that, at points where the fixed-point mapping is differentiable, the fixed-point sensitivity matches the classical KKT-based sensitivity.
\begin{theorem}
\label{thm:fp-kkt}
Suppose that the KKT system of \eqref{eq:mv-multip} satisfies strict complementarity.
Then the Jacobian $\partial \Zbf_t^{*}/\partial \theta$ obtained from the fixed-point formula~\eqref{eq:md-gradient} is identical to the sensitivity derived by differentiating the KKT system.
\end{theorem}
\noindent\textit{Proof.} See~\ref{app:proof-fp-kkt}.

In practice, the matrix inverse in~\eqref{eq:md-gradient} is circumvented by expressing it as a truncated Neumann series.  
Specifically, in our setting, the MDFP update exhibits local contractiveness in a neighborhood of the fixed point.
This property is shown in~\cref{prop:mdfp-spectral-radius} and proved in \ref{app:proof-spectral-radius}.
\begin{proposition}
\label{prop:mdfp-spectral-radius}
Suppose that the KKT system of \eqref{eq:mv-multip} satisfies strict complementarity.
Then, for sufficiently small step size $\eta$, the spectral radius of the Jacobian $\partial_{\Zbf_t^{*}} \Phi(\Zbf_t^{*},\theta)$ is strictly less than $1$.
\end{proposition}
By~\cref{prop:mdfp-spectral-radius}, we obtain
$(I - \partial_{\Zbf_t^{*}}\Phi(\Zbf_t^{*},\theta))^{-1}
=\sum_{b=0}^{\infty}(\partial_{\Zbf_t^{*}}\Phi(\Zbf_t^{*},\theta))^{b}$.
Hence, the gradient can be approximated as
\begin{equation}
\label{eq:neumann-approx}
\frac{\partial \Zbf_t^{*}}{\partial \theta}
\approx
\sum_{b=0}^{B}
\big[\partial_{\Zbf_t^{*}}\Phi(\Zbf_t^{*},\theta)\big]^{b}
\frac{\partial \Phi(\Zbf_t^{*},\theta)}{\partial \theta},
\end{equation}
where $B$ is a small truncation order.
This formulation avoids explicit matrix inversion
and provides a consistent first-order approximation of the exact gradient
while significantly reducing runtime and memory costs during backpropagation.

In our setting, the mirror map is entropic and the update in~\eqref{eq:md-simplex-update} reduces to a softmax-like rescaling on each simplex.
The normalization is performed stage by stage, while the turnover penalty makes the gradient at stage $s$ depend only on the allocations at stages $s-1$, $s$, and $s+1$.
The fixed-point Jacobian therefore inherits a block-tridiagonal structure: each block row contains only the current-stage and adjacent-stage blocks, and all nonadjacent blocks vanish.
The complete blockwise differentiation is provided in the appendix proof of~\cref{thm:fp-kkt}.
This local structure allows each Neumann term in~\eqref{eq:neumann-approx} to be evaluated through inexpensive block-banded Jacobian-vector products that exchange information only between adjacent stages and reuse the same elementwise multiplications and normalizations as the forward MD map.
Consequently, the fixed-point implicit gradient matches the sensitivity of KKT-based differentiation without explicitly forming or factorizing large KKT matrices.

\subsection{Framework architecture}

\begin{algorithm}[t]
\caption{\ee: integrated multi-period optimization}
\label{alg:tempo-admm-md}
\begin{algorithmic}

\Require Training data $\{(X_s, Y_s), s=t-T-H+1,\dots,t-H\}$, 
risk parameter $\delta$, penalty $\lambda$, smoothing factor $\kappa$, stepsize $\eta$, small truncation order $B$
\Ensure First-step allocation $z^*_{t+1}$
\State Initialize parameters $\theta$;
\For{each training step}
    \For{$s = t-T-H+1,\ldots,t-H$}
        \State Predict future returns 
               $\hat{Y}_s=\phi_\theta(X_s)$;
        \State Estimate covariance $\hat{V}_{k},\ k=s+1,\ldots,s+H$;
        \State Solve the multi-period problem \eqref{eq:mv-multip};
        \State Compute loss contribution 
               $\ell_{\mathrm{d}}(\Zbf^{*}_s(\theta), Y_s)$;
        \State Compute $\frac{\partial \Zbf^{*}_s(\theta)}{\partial\theta}$ using \eqref{eq:neumann-approx};    
    \EndFor
    \State Update $\theta$ using back-propagation with $\nabla_\theta \mathcal{L}_{\mathrm{d}}(\theta)$ in \eqref{eq:bw-chain};
\EndFor

\State Compute $\hat{Y}_t=\phi_{\theta}(X_t)$;
\State Compute $\hat{V}_s,\ s=t+1,\ldots,t+H$ based on past returns;
\State Solve \eqref{eq:mv-multip} to obtain  $\Zbf_t^*$ and execute the first-step portfolio $z^*_{t+1}$.

\end{algorithmic}
\end{algorithm}

\begin{figure*}
  \centering
  \includegraphics[width=\textwidth]{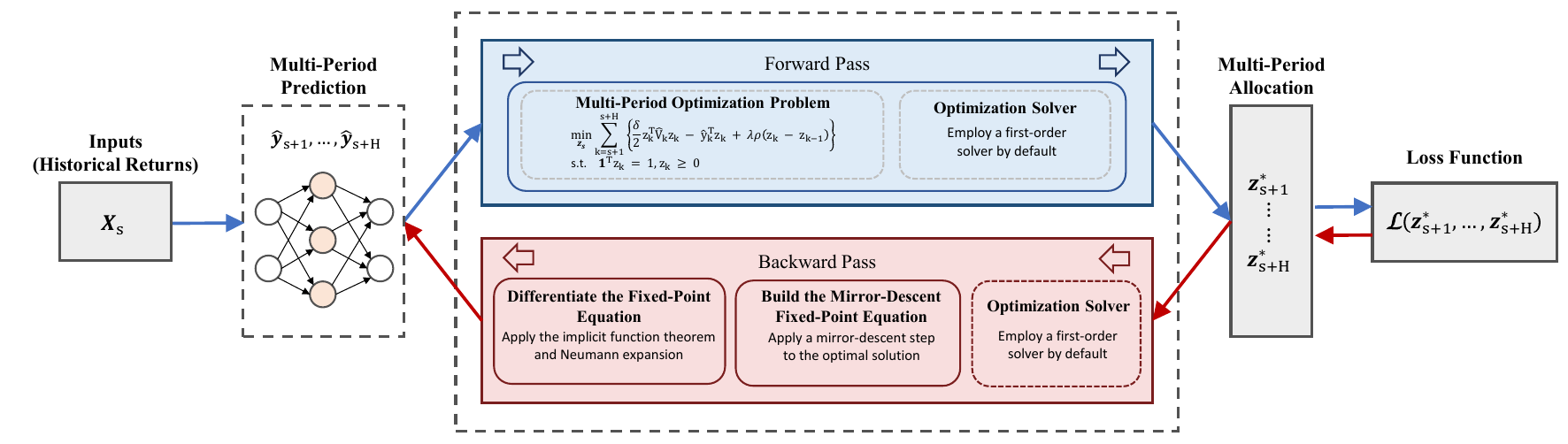}
  \caption{Overview of IPMO. The predictor generates multi-period return predictions, which are passed to a differentiable optimization layer to produce the portfolio allocations. The fixed-point formulation enables implicit differentiation through the solver, allowing the entire multi-period decision process to be trained jointly.
  }
  \label{fig:workflow}
\end{figure*}

The \ee\ framework integrates  optimization and learning into a single differentiable pipeline.
Given a rolling window of past returns, the prediction model $\phi_\theta$ generates per-stage optimization parameters. The inner multi-stage allocation problem is solved on the simplex of portfolio weights with a smoothed objective (here we use MOSEK \citep{andersen2000mosek}), and the resulting multi-period allocations are evaluated by a downstream decision loss that guides the parameter updates.
The overall procedure is summarized by \cref{alg:tempo-admm-md} and visualized by \cref{fig:workflow}. The forward pass computes optimal allocations through the inner solver, while the backward pass propagates gradients via the mirror-descent fixed-point (MDFP) formulation.

\subsubsection{Forward pass}

The forward pass maps a window of past returns to multi-period allocations by solving the smoothed allocation problem.  
For each training date $s \in \{t-T-H+1, \dots, t-H\}$, the predictor $\phi_\theta$ produces multi-period forecasts of returns $\hat{Y}_s$.
Given these parameters and the pre-trade allocation, 
the inner optimization follows \eqref{eq:mv-multip}.
The optimizer directly solves this convex program to obtain
$\Zbf_s^{*}(\theta)
=\arg\min_{\Zbf_s\in\boldsymbol{\Omega}}\hat{F}_s(\Zbf_s,\hat{Y}_s)$.
The resulting allocation $\Zbf_s^{*}(\theta)$ defines the forward decision output for date $s$,  
which is evaluated by the downstream decision loss in~\eqref{eq:tempo-mv-final}.  
Consistent with the MPC approach, the network outputs an entire sequence of multi-period portfolio weights, while only the first-step allocation is executed.
All solver hyperparameters are kept unchanged between training and testing.

\subsubsection{Backward pass}

The backward pass propagates gradients from the evaluation loss back to the predictor parameters.  
The detached gradient of the empirical loss with respect to~$\theta$ follows the chain rule:
\begin{equation}
\label{eq:bw-chain}
\nabla_\theta \mathcal{L}_{\mathrm{d}}(\theta)
=
\frac{1}{T}\sum_{s = t-T-H+1}^{t-H}\frac{\partial \ell_{\mathrm{d}}\big(\Zbf_s^{*}(\theta),Y_s\big)}{\partial \Zbf_s^{*}(\theta)}
\frac{\partial \Zbf_s^{*}(\theta)}{\partial \theta}.
\end{equation}
The first partial derivative represents the sensitivity of the decision loss to the allocation, with $Y_s$ held fixed.
The second partial derivative differentiates through the inner allocation program, whose solution $\Zbf_s^{*}(\theta)$ is defined by the MDFP layer~\eqref{eq:md-fp} and its implicit gradient expression~\eqref{eq:md-gradient}. The parameters are updated based on the average gradient of the training batch.

\section{Experiments}

\label{section:exp}
To demonstrate the effectiveness of the \ee\ framework and the proposed algorithm, we present computational results on real market data. Our analysis is divided into two parts. The first part compares outcomes of the \ee\ model with the two-stage model under the MPC setting. We connect the portfolio behavior to the underlying models by analyzing the dynamics of portfolio weights and rolling predictions. The second part evaluates the efficiency of different end-to-end learning frameworks. We benchmark our differentiable optimization layer against the most commonly used methods in differentiable optimization.

\subsection{Data and computational setup}
We use daily asset returns for seven exchange-traded funds (ETFs) of major asset classes: 
VTI (Vanguard Total Stock Market ETF), IWM (iShares Russell 2000 ETF), AGG (iShares Core U.S. Aggregate Bond ETF),
LQD (iShares iBoxx Investment Grade Corporate Bond ETF), MUB (iShares National Muni Bond ETF),
DBC (Invesco DB Commodity Index Tracking Fund), and GLD (SPDR Gold Shares)
over the time period 2011-2024, following \citet{Uysal2024}. We keep the first eight years (2011-2018/12) for hyperparameter tuning and the remaining six years (2019-2024) for out-of-sample testing.

We consider two standard neural networks commonly used in multi-period time series forecasting and return prediction: the RLinear model~\citep{li2023revisiting} with L2 regularization and the CNN-LSTM model~\citep{MURRAY2024103791}. We experiment with both simple and sophisticated models to demonstrate the performance gains of end-to-end learning under different model complexities. In the RLinear model, each asset return series is independently modeled by a single fully connected layer, and we average historical returns over five-day intervals to mitigate estimation noise. The CNN-LSTM model applies a pooled model for all assets, as in \citet{MURRAY2024103791}. The CNN-LSTM model uses a Convolutional Neural Network (CNN) with 64 channels, a kernel size of 5, a stride of 1, and the ReLU activation. The CNN output is passed into a max-pooling layer with a kernel size of 2. The Long Short-Term Memory (LSTM) model has two layers with a hidden dimension of 64 and generates the final predictions. Both neural networks incorporate reversible instance normalization to address the distribution shift problem~\citep{kim2022reversible}. The predictive models are trained to predict the returns for the next $H$ days using historical returns from the past 120 days and are re-trained every 20 days using a look-back window of 250 days. To keep the estimation process simple, the covariance estimation follows a simple exponentially weighted moving average (EWMA) method, updated daily using the historical returns of the past 20 days. The portfolio is rebalanced daily. On each day, the model generates portfolio weights over the prediction horizon, but only the allocation for the next period is applied, following \citet{boyd2017}. The hyperparameters include the learning rate $\gamma\in\{0.001, 0.002, 0.005, 0.01\}$, the turnover penalty $\lambda\in\{0.0001, 0.0005, 0.001, 0.005, 0.01\}$, and the planning horizon $H\in\{1, 5, 10, 20, 50\}$. In addition, we tune the L2 regularizer for the linear model over the range $\{0, 0.0001, 0.001\}$. For both the two-stage (TS) and \ee\ frameworks, we select the hyperparameters based on the in-sample Sharpe ratio and turnover rate. Among the ten hyperparameter configurations with the highest Sharpe ratios, we choose the one with the lowest turnover.
\revise{For the forward optimization, MOSEK's conic interior-point tolerances for primal feasibility, dual feasibility, the relative optimality gap, infeasibility detection, and complementarity reduction are all set to $10^{-6}$. We fix the MDFP step size at $\eta=10$.}

We benchmark our \ee\ framework against standard portfolio allocation baselines, including an equal-weighted (EW) portfolio and a classical mean-variance (MV) portfolio. The EW portfolio uses constant $1/N$ weights for all $N$ assets, rebalanced daily. The MV portfolio is constructed using the sample mean of the past 120-day returns and the same covariance estimators as above. Each predictive model is evaluated within the two-stage and the \ee\ frameworks. We report results both before and after accounting for \revise{2\,bps} one-way transaction costs. \revise{This benchmark is motivated by the liquidity evidence in \citet{khomyn2024value}.}
All experiments were conducted on a machine with an Intel\textsuperscript{\textregistered} Core\textsuperscript{TM} Ultra~9~285H CPU and an NVIDIA GeForce RTX~5070 Laptop GPU (8\,GB), using PyTorch~2.2 with CUDA~12.4.

\subsection{Comparison of performance}
\begin{figure*}
\centering
\includegraphics[width=0.9\textwidth]{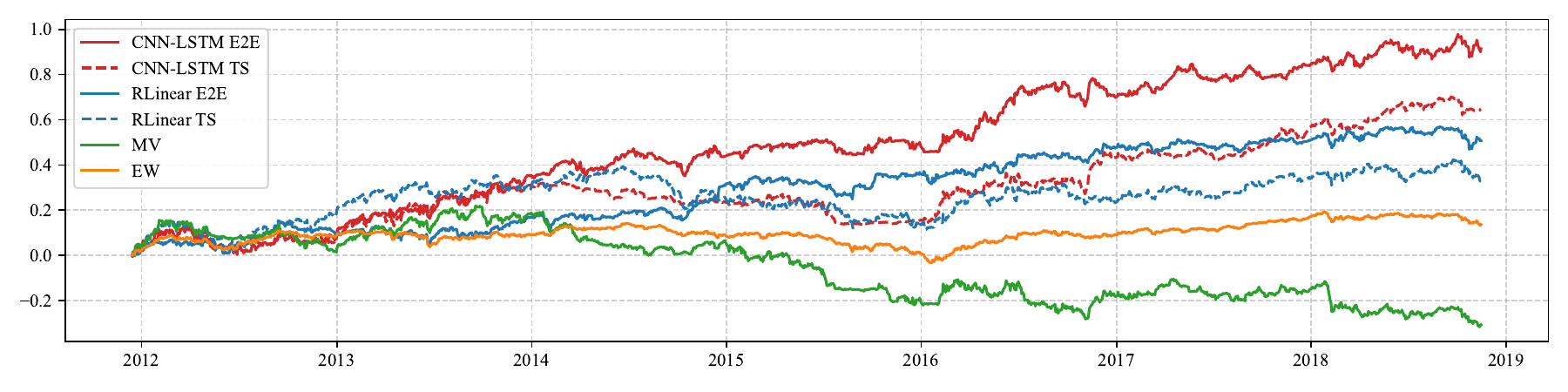}
\includegraphics[width=0.9\textwidth]{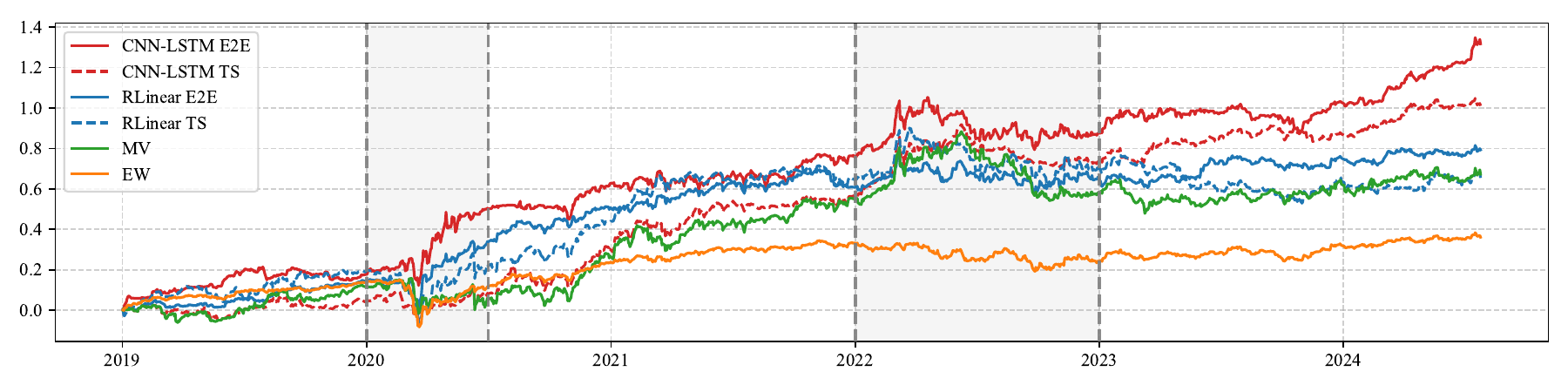}
\caption{Optimal portfolio cumulative log returns net of \revise{2\,bps} transaction costs for six strategies over in-sample (top) and out-of-sample (bottom) periods. The gray shaded regions in the out-of-sample panel indicate the first half of 2020 and the full year 2022.}
\label{fig:nav-all}
\end{figure*}

\begin{table*}[h!]
\centering
\scriptsize
\caption{Annualized portfolio performance statistics gross and net of \revise{2\,bps} transaction costs over in-sample (2011-2018) and out-of-sample (2019-2024) periods.}
\label{tab:vertical-results}
\setlength{\tabcolsep}{6pt}
\renewcommand{\arraystretch}{1.1}
\begin{tabular}{lrrrrrrr}
\toprule
Portfolio & Return & Volatility & Sharpe & MDD & Calmar & Return/ave.DD & Turnover \\
\midrule\multicolumn{8}{l}{\textbf{2011-2018/12 Gross of transaction costs}}\\

CNN-LSTM \ee\ &  0.1451 &      0.1329 &  \textbf{1.0922} & 0.1199 &  \textbf{1.2100} &           4.6451 &    0.0850 \\
 CNN-LSTM TS &  0.1047 &      0.1128 &  0.9279 & 0.1627 &  0.6436 &           2.4530 &    0.0967 \\
RLinear \ee\ &  0.1021 &      0.0983 &  1.0383 & 0.0931 &  1.0964 &           \textbf{5.9420} &    0.4719 \\
 RLinear TS &  0.0898 &      0.1219 &  0.7362 & 0.2031 &  0.4420 &           1.7808 &    0.6923 \\
         EW &  0.0217 &      0.0596 &  0.3640 & 0.1604 &  0.1353 &           0.5969 &    0.0028 \\
         MV & -0.0242 &      0.1246 & -0.1940 & 0.3745 & -0.0646 &          -0.1229 &    0.2513 \\

\midrule\multicolumn{8}{l}{\textbf{2011-2018/12 Net of transaction costs}}\\

CNN-LSTM \ee\ &  0.1411 &      0.1329 &  \textbf{1.0623} & 0.1199 &  \textbf{1.1767} &           \textbf{4.5590} &    0.0850 \\
 CNN-LSTM TS &  0.1000 &      0.1128 &  0.8860 & 0.1722 &  0.5806 &           2.1984 &    0.0967 \\
RLinear \ee\ &  0.0783 &      0.0983 &  0.7965 & 0.0931 &  0.8412 &           4.5172 &    0.4719 \\
 RLinear TS &  0.0550 &      0.1218 &  0.4516 & 0.2423 &  0.2271 &           0.7071 &    0.6922 \\
         EW &  0.0215 &      0.0596 &  0.3603 & 0.1607 &  0.1336 &           0.5874 &    0.0028 \\
         MV & -0.0366 &      0.1246 & -0.2938 & 0.4144 & -0.0884 &          -0.1714 &    0.2513 \\

\midrule\multicolumn{8}{l}{\textbf{2019-2024/12 Gross of transaction costs}}\\

CNN-LSTM \ee\ &  0.2591 &      0.1921 &  \textbf{1.3490} & 0.2271 &  \textbf{1.1410} &           \textbf{5.5592} &    0.0740 \\
 CNN-LSTM TS &  0.1985 &      0.1566 &  1.2673 & 0.1881 &  1.0552 &           4.1896 &    0.0862 \\
RLinear \ee\ &  0.1815 &      0.1463 &  1.2404 & 0.1668 &  1.0881 &           5.0095 &    0.5543 \\
 RLinear TS &  0.1730 &      0.1919 &  0.9017 & 0.2660 &  0.6506 &           1.9801 &    0.7351 \\
         EW &  0.0705 &      0.1008 &  0.6994 & 0.2073 &  0.3400 &           2.0793 &    0.0043 \\
         MV &  0.1479 &      0.1898 &  0.7794 & 0.3318 &  0.4459 &           1.3533 &    0.1797 \\

\midrule\multicolumn{8}{l}{\textbf{2019-2024/12 Net of transaction costs}}\\

CNN-LSTM \ee\ &  0.2558 &      0.1920 &  \textbf{1.3319} & 0.2271 &  \textbf{1.1264} &           \textbf{5.4880} &    0.0740 \\
 CNN-LSTM TS &  0.1943 &      0.1566 &  1.2407 & 0.1893 &  1.0264 &           4.0020 &    0.0862 \\
RLinear \ee\ &  0.1536 &      0.1463 &  1.0500 & 0.1668 &  0.9208 &           4.2395 &    0.5543 \\
 RLinear TS &  0.1363 &      0.1919 &  0.7101 & 0.3105 &  0.4390 &           1.3159 &    0.7349 \\
         EW &  0.0702 &      0.1008 &  0.6966 & 0.2073 &  0.3385 &           2.0612 &    0.0043 \\
         MV &  0.1392 &      0.1898 &  0.7337 & 0.3324 &  0.4189 &           1.2442 &    0.1797 \\
\bottomrule
\end{tabular}
\end{table*}
\cref{fig:nav-all} shows cumulative returns net of transaction costs for in-sample and out-of-sample periods, and performance statistics are reported in \cref{tab:vertical-results}.
\revise{The CNN-LSTM strategies outperform EW and MV in full-period net return and Sharpe ratio in both samples. Relative to two-stage training, \ee\ achieves higher full-period net returns and Sharpe ratios and lower turnover under both predictors in both samples.}

In the in-sample period, CNN-LSTM \ee\ achieves the highest net return of 0.1411, Sharpe ratio of 1.0623, Calmar ratio of 1.1767, and return to average drawdown ratio of 4.5590. Relative to CNN-LSTM TS, it earns a higher net return (0.1411 versus 0.1000) and Sharpe ratio (1.0623 versus 0.8860), and reduces maximum drawdown from 0.1722 to 0.1199. Its volatility is 0.1329, compared with 0.1128 for TS. Under RLinear, \ee\ increases the annualized net return from 0.0550 to 0.0783 and the Sharpe ratio from 0.4516 to 0.7965, while reducing maximum drawdown from 0.2423 to 0.0931. The Calmar and return to average drawdown ratios also increase (0.8412 versus 0.2271, and 4.5172 versus 0.7071). EW earns a net return of 0.0215 with a Sharpe ratio of 0.3603, while MV records a net return of -0.0366 and a Sharpe ratio of -0.2938.

In the out-of-sample period, CNN-LSTM \ee\ achieves the highest annualized net return of 0.2558, Sharpe ratio of 1.3319, Calmar ratio of 1.1264, and return to average drawdown ratio of 5.4880. Compared with CNN-LSTM TS, it earns a higher net return (0.2558 versus 0.1943) and Sharpe ratio (1.3319 versus 1.2407). \revise{However, its volatility (0.1920 versus 0.1566) and maximum drawdown (0.2271 versus 0.1893) are also higher, so these return and Sharpe improvements do not imply lower risk across all metrics.} Under RLinear, \ee\ increases the net return from 0.1363 to 0.1536 and the Sharpe ratio from 0.7101 to 1.0500, while lowering maximum drawdown from 0.3105 to 0.1668. EW and MV earn net returns of 0.0702 and 0.1392, with Sharpe ratios of 0.6966 and 0.7337, respectively. RLinear \ee\ has a higher Sharpe ratio than both benchmarks, while RLinear TS slightly underperforms MV in Sharpe ratio.

Furthermore, the \ee\ model \revise{has lower turnover than the two-stage model in both periods. Out of sample, turnover falls from 0.0862 to 0.0740 for CNN-LSTM and from 0.7349 to 0.5543 for RLinear, reductions of approximately 14.2\% and 24.6\%, respectively. The corresponding reduction in the cost burden is especially material for RLinear.}
During the in-sample period, transaction costs reduce annualized returns by 0.40 and 0.47 percentage points for CNN-LSTM \ee\ and TS, respectively. With RLinear, the reductions are 2.38 and 3.48 percentage points. During the out-of-sample period, the corresponding reductions are 0.33 and 0.42 percentage points for CNN-LSTM, and 2.79 and 3.67 percentage points for RLinear. \revise{Under both predictors, \ee\ incurs a smaller cost deduction than two-stage training and retains higher full-period net returns and Sharpe ratios.}

\begin{figure*}[h!]
  \centering
  \includegraphics[width=0.9\textwidth]{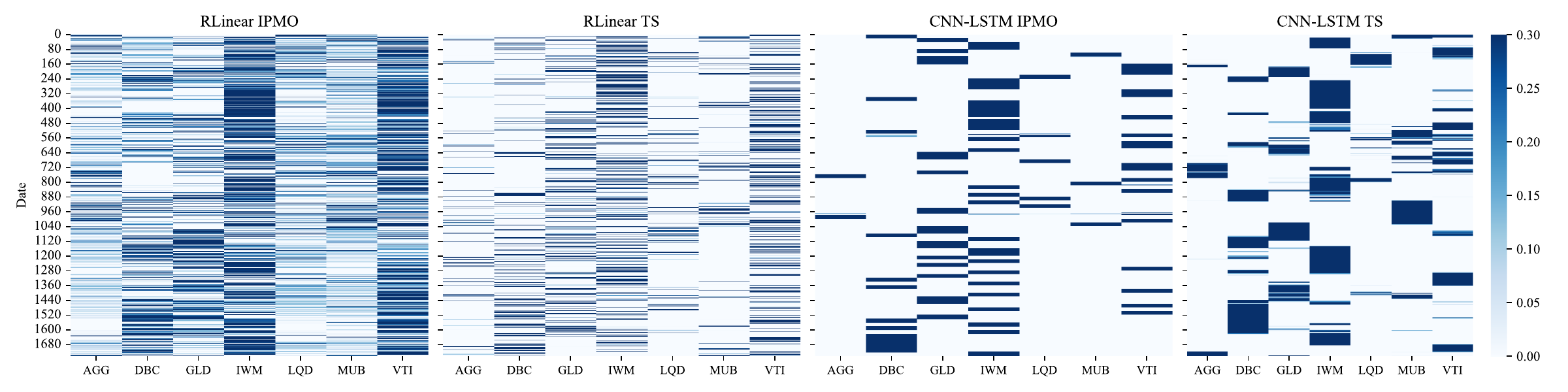}
  \includegraphics[width=0.9\textwidth]{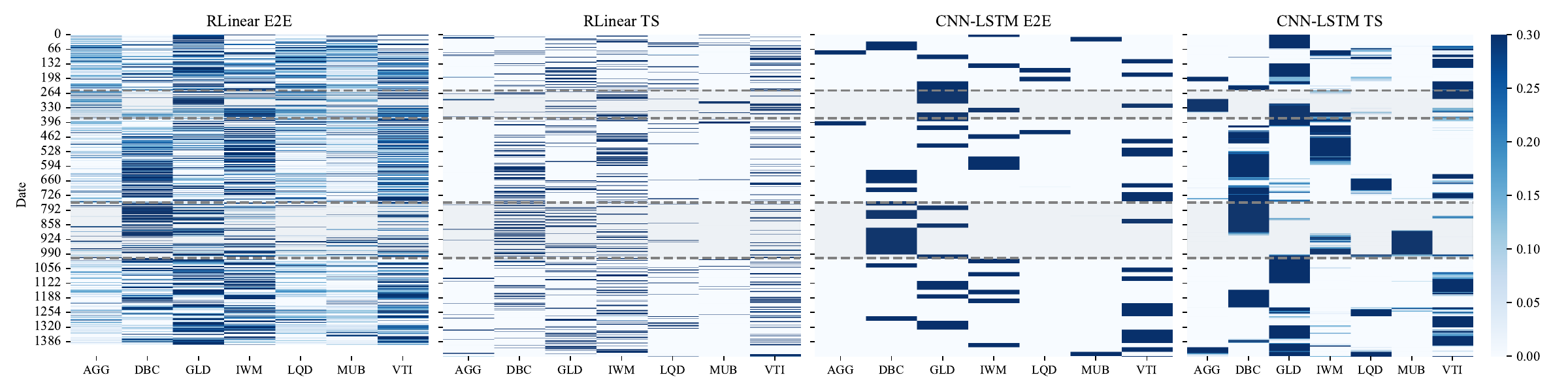}
  \caption{Optimal portfolio weights across seven ETFs for four strategies over in-sample (top) and out-of-sample (bottom) periods. The gray dashed boundaries in the out-of-sample panel delimit the first half of 2020 and the full year 2022.}
  \label{fig:weights}
\end{figure*}

Next, we visualize the portfolio weights over time using a heatmap in \cref{fig:weights}. For RLinear, the two-stage model tends to hold large positions across several assets and frequently makes abrupt shifts in the set of holdings.
Instead, the RLinear \ee\ model exhibits smoother trajectories, where weight adjustments occur in smaller increments and remain stable over longer intervals. 
For CNN-LSTM, both models maintain highly concentrated positions, holding a single asset for extended periods.
Nonetheless, the two-stage model still exhibits instantaneous reallocations occasionally, while the \ee\ model shows a buy-and-hold behavior and the reallocation frequency is more consistent.

The shaded intervals in \cref{fig:nav-all,fig:weights} isolate two out-of-sample stress episodes with distinct economic origins: the COVID-19 shock in the first half of 2020 and the monetary-tightening regime of 2022. \Cref{tab:stress-2020-h1,tab:stress-2022} report annualized performance net of transaction costs for these periods.
During the first half of 2020, CNN-LSTM \ee\ followed a persistent rotation among GLD, VTI, and IWM before returning to GLD. It achieved an annualized net return of 0.7093 and a Sharpe ratio of 2.2292, versus 0.0966 and 0.5589 for TS, while reducing maximum drawdown from 0.1466 to 0.1253. RLinear \ee\ also outperformed TS, with a net return of 0.4156 versus 0.1225, a Sharpe ratio of 1.9871 versus 0.3830, and maximum drawdown of 0.1301 versus 0.2312. The COVID-19 shock amplified uncertainty and the economic cost of decision-relevant forecast errors \citep{altig2020economic}. By aligning forecasts with the downstream risk-return and turnover objective, \ee\ could prioritize economically material errors. This interpretation is consistent with theoretical evidence that integrated estimation-optimization can be advantageous under model misspecification \citep{elmachtoub2023estimate,elmachtoub2025dissecting}.

\begin{table}[!htbp]
\centering
\scriptsize
\caption{Performance net of \revise{2\,bps} transaction costs during the first half of 2020. Return and volatility are annualized.}
\label{tab:stress-2020-h1}
\setlength{\tabcolsep}{5pt}
\renewcommand{\arraystretch}{1.08}
\begin{tabular}{lrrrrrr}
\toprule
Portfolio & Return & Volatility & Sharpe & MDD & Calmar & Turnover \\
\midrule
CNN-LSTM \ee\ & 0.7093 & 0.3182 & 2.2292 & 0.1253 & 5.6615 & 0.0623 \\
CNN-LSTM TS   & 0.0966 & 0.1728 & 0.5589 & 0.1466 & 0.6589 & 0.0647 \\
RLinear \ee\  & 0.4156 & 0.2091 & 1.9871 & 0.1301 & 3.1939 & 0.5596 \\
RLinear TS    & 0.1225 & 0.3198 & 0.3830 & 0.2312 & 0.5297 & 0.7929 \\
MV            & -0.0835 & 0.2510 & -0.3326 & 0.1596 & -0.5230 & 0.0765 \\
EW            & -0.0155 & 0.2128 & -0.0731 & 0.2073 & -0.0750 & 0.0084 \\
\bottomrule
\end{tabular}
\end{table}

The 2022 episode differs because rapid interest-rate increases altered prevailing cross-asset trends, consistent with the documented effects of monetary-policy shocks on asset prices \citep{rigobon2004impact}. CNN-LSTM TS captured part of this shift through sustained DBC exposure followed by MUB, earning 0.2000 versus 0.1524 for CNN-LSTM \ee\revise{, with a higher Sharpe ratio (0.9655 versus 0.5920), lower volatility (0.2071 versus 0.2574), and lower maximum drawdown (0.1893 versus 0.2277)}. CNN-LSTM \ee\ had slightly lower turnover (0.0600 versus 0.0615) and average total variation (0.0485 versus 0.0562). \revise{These measures describe allocation stability and do not establish superior risk control in this episode.} Under RLinear, \ee\ earned 0.0741 versus 0.0891 for TS, with lower volatility (0.2056 versus 0.2508), maximum drawdown (0.1809 versus 0.2512), and turnover (0.6232 versus 0.6738). The corresponding Sharpe ratios were 0.3603 and 0.3555. \revise{Thus, the full-period improvements in net return and Sharpe ratio do not extend to every market regime.}

\begin{table}[!htbp]
\centering
\scriptsize
\caption{Performance net of \revise{2\,bps} transaction costs during 2022. Return and volatility are annualized.}
\label{tab:stress-2022}
\setlength{\tabcolsep}{5pt}
\renewcommand{\arraystretch}{1.08}
\begin{tabular}{lrrrrrr}
\toprule
Portfolio & Return & Volatility & Sharpe & MDD & Calmar & Turnover \\
\midrule
CNN-LSTM \ee\ & 0.1524 & 0.2574 & 0.5920 & 0.2277 & 0.6692 & 0.0600 \\
CNN-LSTM TS   & 0.2000 & 0.2071 & 0.9655 & 0.1893 & 1.0561 & 0.0615 \\
RLinear \ee\  & 0.0741 & 0.2056 & 0.3603 & 0.1809 & 0.4095 & 0.6232 \\
RLinear TS    & 0.0891 & 0.2508 & 0.3555 & 0.2512 & 0.3548 & 0.6738 \\
MV            & 0.0630 & 0.2444 & 0.2577 & 0.2884 & 0.2184 & 0.0822 \\
EW            & -0.0828 & 0.1182 & -0.7000 & 0.1302 & -0.6355 & 0.0058 \\
\bottomrule
\end{tabular}
\end{table}

\begin{figure}[h!]
  \centering
  \begin{subfigure}[t]{0.48\textwidth}
    \centering
    \includegraphics[width=\linewidth]{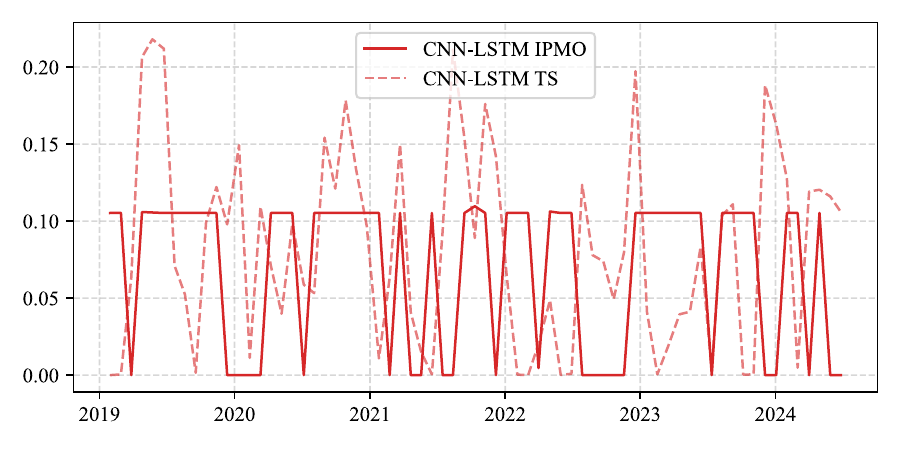}
  \end{subfigure}\hfill
  \begin{subfigure}[t]{0.48\textwidth}
    \centering
    \includegraphics[width=\linewidth]{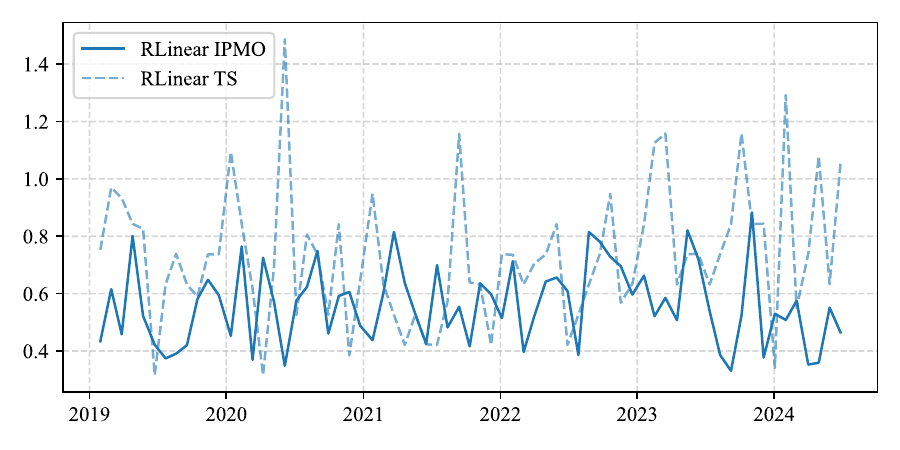}
  \end{subfigure}
  \caption{Average total variation of out-of-sample portfolio weights for four strategies over non-overlapping 20-day windows.}
  \label{fig:tv}
\end{figure}

\cref{fig:tv} reports the rolling 20-day total variation (TV) of portfolio weights.
The difference in allocation behavior observed in \cref{fig:weights} is clearly reflected in their TV dynamics. For RLinear, the two-stage model exhibits both higher peak values and stronger fluctuations, implying large and frequent reallocations among assets. In contrast, the \ee\ model keeps TV within a consistently lower and narrower range, indicating smoother transitions and gradual, small-magnitude adjustments over time.
For CNN-LSTM, both models achieve much lower overall TV levels, consistent with their concentrated and long-holding allocation patterns. Yet, the two-stage model still displays occasional surges in TV, suggesting abrupt switches, while the \ee\ model's TV remains organized around two discrete levels, approximately 0 and 0.105, corresponding to long stationary phases punctuated by short, well-defined rebalancing events. This discrete switching pattern signifies a highly organized low-frequency turnover regime.
Overall, these results confirm that the \ee\ model delivers more temporally stable and coherent allocation patterns across both predictor types.

\begin{figure*}
  \centering
  \includegraphics[width=0.8\textwidth]{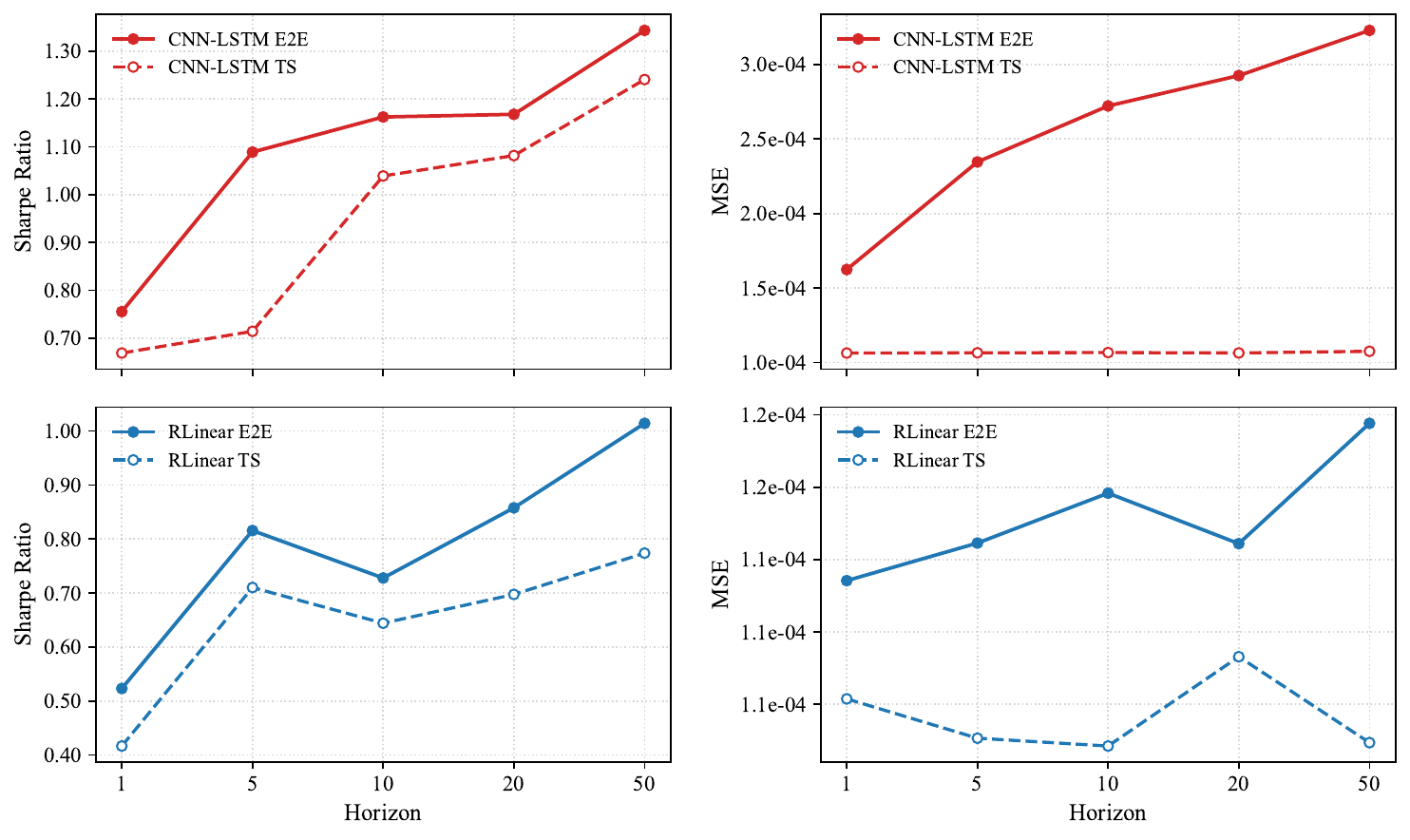}
  \caption{Sharpe ratios \revise{net of 2\,bps transaction costs} (left) and MSE (right) across forecasting horizons for CNN-LSTM (top) and RLinear (bottom) models.}
  \label{fig:h_sensitivity}
\end{figure*}

To evaluate whether the performance advantage of the \ee\ model persists across different planning horizons and whether this advantage stems from improved accuracy of return predictions, \cref{fig:h_sensitivity} jointly reports the out-of-sample Sharpe ratio and mean squared forecast error (MSE) for each $H$.
Across all configurations, the \ee\ model consistently attains higher Sharpe ratios than the two-stage benchmark.
For both the \ee\ model and the two-stage model, Sharpe ratios generally improve with longer horizons and reach their highest values when $H{=}50$.
It is worth noting that when $H{=}1$, corresponding to a single-period formulation, all methods deliver their lowest Sharpe ratio. This degradation reflects the two limitations discussed earlier:
the inability of a single-stage objective to internalize the intertemporal impact of transaction costs and the absence of multi-period coupling, which leads to inconsistent forecasts and conflicting signals.
Examining model-specific patterns, the RLinear predictor displays a modest dip in Sharpe ratio around $H{=}10$, after which the gap shifts decisively in favor of the \ee\ model.
For the higher-capacity CNN-LSTM predictor, performance improves almost monotonically with $H$, and the \ee\ model maintains a clear lead throughout.
The persistence of this gap across predictors indicates that the advantage of \ee\ is robust.

The MSE results further clarify the source of these gains in Sharpe ratio.
Under RLinear, the \ee\ model attains slightly higher MSE than the two-stage model across all horizons, yet consistently achieves superior Sharpe ratios.
This pattern indicates that the \ee\ model sacrifices a small amount of pointwise predictive accuracy in exchange for reduced decision error, producing forecasts that align more closely with the multi-period, transaction-cost-aware objective.
For CNN-LSTM, the two-stage model shows the lowest and almost horizon-invariant MSE, suggesting that it has reached the accuracy limit achievable by the underlying architecture, which directly contributes to its performance improvement.
In contrast, the \ee\ model produces the largest MSE, and this discrepancy grows systematically with $H$.
This reflects that, with a higher-capacity prediction model such as CNN-LSTM, a larger planning window enables the end-to-end objective to induce forecast adjustments that depart more substantially from pointwise accuracy in order to encode richer intertemporal structure relevant for the multi-period decision problem.
Despite these larger deviations, the \ee\ model still attains the best post-fee performance, indicating that the resulting biases are decision-useful and shape the forecast path in a way that more effectively supports multi-period allocation.

We assess training stability by independently retraining each predictor under \ee\ and TS over 100 random seeds with fixed selected hyperparameters. \revise{Returns are net of 2\,bps one-way transaction costs.} For each predictor, we conduct a one-sided Welch \(t\)-test of
\[
H_0:\mu_{\ee}\leq\mu_{\mathrm{TS}}
\qquad\text{versus}\qquad
H_1:\mu_{\ee}>\mu_{\mathrm{TS}},
\]
\revise{where each mean is taken over training randomness conditional on the evaluated market history.} For RLinear, the mean annualized net return (sample standard deviation) is \(0.1496\) (\(0.0621\)) for \ee\ and \(0.1309\) (\(0.0392\)) for TS, with \(t=2.55\) and one-sided \(p=0.0059\). For CNN-LSTM, the corresponding values are \(0.2508\) (\(0.0777\)) and \(0.1937\) (\(0.0588\)), with \(t=5.86\) and one-sided \(p<0.0001\). Both tests reject the null at the 1\% level. \revise{These results assess variation over training randomness on the observed market histories and do not establish significance with respect to uncertainty in the market sample.}

\revise{Separately, we assess market-sample uncertainty in the difference between the two methods' performance averaged over the 100 runs, using a paired studentized circular block bootstrap following \citet{ledoit2008robust}. We align common out-of-sample dates within each predictor and jointly resample 20-day blocks across both methods and all runs, with 10,000 resamples. Each run's net return and Sharpe ratio are recomputed before averaging within method. The 95\% confidence intervals in \cref{tab:seed-paired-inference} exclude zero; this conclusion also holds for block lengths of 5, 10, 40, and 60 days. This inference conditions on the fitted runs and assumes weak temporal dependence and approximate stationarity of the out-of-sample return process.}

\begin{table}[!htbp]
\centering
\small
\caption{\revise{Paired block-bootstrap inference for differences in out-of-sample performance averaged over 100 runs (\ee\ minus TS), net of 2\,bps transaction costs. $\Delta R$ is the annualized net-return difference and $\Delta\mathrm{SR}$ is the difference in mean annualized Sharpe ratios. Intervals use 20-day blocks and 10,000 resamples.}}
\label{tab:seed-paired-inference}
\setlength{\tabcolsep}{5pt}
\renewcommand{\arraystretch}{1.08}
\revise{
\begin{tabular}{lrrrr}
\toprule
Predictor & $\Delta R$ & 95\% CI & $\Delta\mathrm{SR}$ & 95\% CI \\
\midrule
RLinear & 0.0186 & [0.0029, 0.0343] & 0.1621 & [0.0648, 0.2593] \\
CNN-LSTM & 0.0571 & [0.0397, 0.0745] & 0.3304 & [0.2151, 0.4457] \\
\bottomrule
\end{tabular}}
\end{table}

\subsection{Large-universe experiment with weekly rebalancing}
\label{sec:crsp-weekly}

To assess performance beyond the seven-ETF setting, we consider a dynamic universe of 100 U.S. stocks from CRSP with weekly rebalancing. \revise{Every four weeks, we select the 100 eligible stocks with the largest average market capitalization over the preceding 250 trading days.} Stocks must be active and have sufficient return history for estimation. Portfolios are rebalanced at each week's final trading close, normally Friday. We consider planning horizons of \(H\in\{1,2,3,4,5,10,20\}\) weeks and implement only the first allocation before reoptimizing the following week.

\begin{figure}[H]
\singlespacing
\centering
\includegraphics[width=0.92\textwidth]{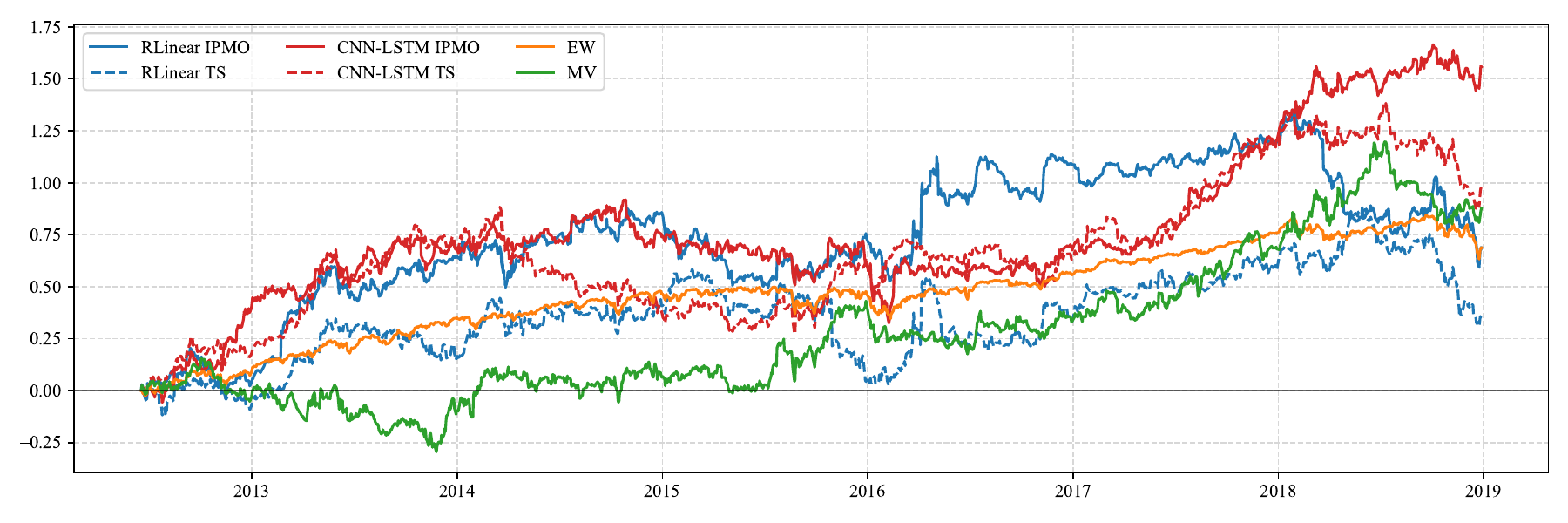}
\includegraphics[width=0.92\textwidth]{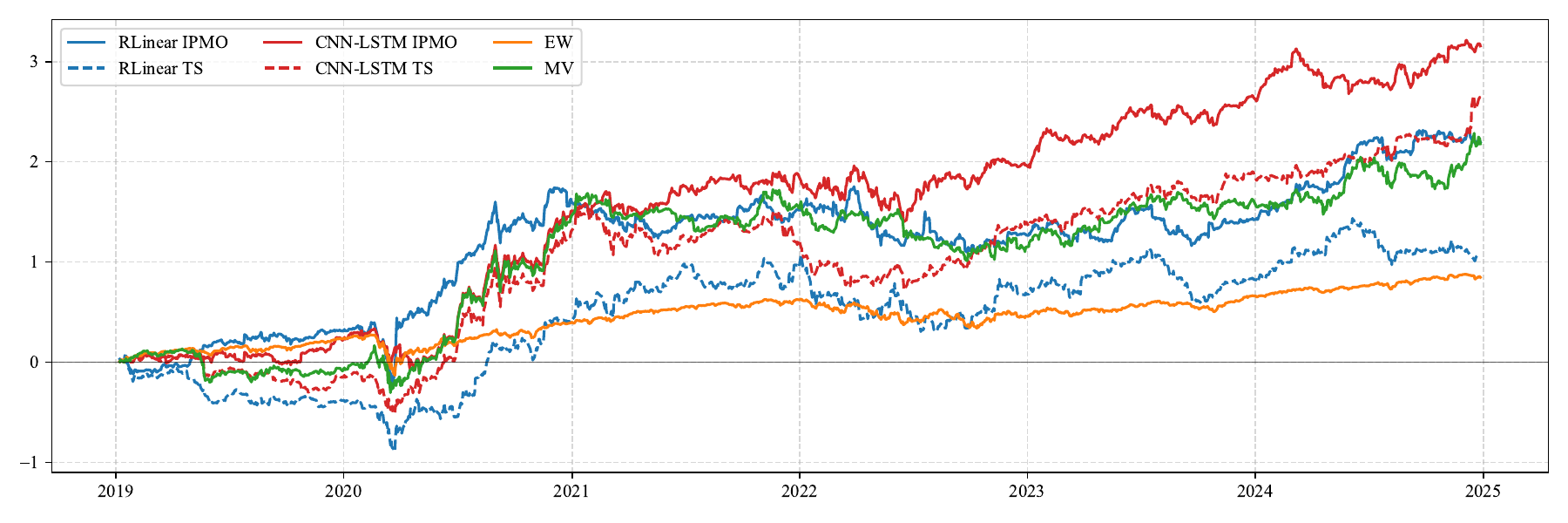}
\caption{Optimal portfolio cumulative log returns net of transaction costs for six weekly rebalanced strategies in the 100-stock CRSP experiment over the in-sample (top) and out-of-sample (bottom) periods.}
\label{fig:crsp-weekly-nav}
\end{figure}

Predictors use the preceding 120 daily returns to forecast compounded weekly returns and are retrained every four weeks on a rolling window of 50 weekly samples. Covariances are updated by EWMA using rolling 20-day windows \revise{and held fixed across each planning horizon. We use a covariance regularization coefficient of $\varepsilon=10^{-6}$ in daily-return units and a risk coefficient of $1$ on the daily covariance scale, equivalent to $0.2$ under a five-trading-day weekly risk approximation.} Hyperparameters are selected over 2011-2018 using the same rule as in the ETF experiment, and 2019-2024 is reserved for out-of-sample evaluation.

We report gross and net performance with 20\,bps \revise{one-way} costs applied to the full $L_1$ traded amount\revise{, following \citet{kan2022optimal}}. \revise{Holdings are tracked by PERMNO and drift with returns between weekly rebalances. Stocks leaving the universe are liquidated at rebalancing, with their sales included in transaction costs; no initial entry charge is applied.}

\begin{table}[H]
\singlespacing
\centering
\scriptsize
\caption{Portfolio performance in the 100-stock CRSP experiment, gross and net of transaction costs, over the in-sample (2011-2018) and out-of-sample (2019-2024) periods. Return and volatility are annualized; turnover is the average at weekly rebalancing dates.}
\label{tab:crsp-weekly-results}
\setlength{\tabcolsep}{6pt}
\renewcommand{\arraystretch}{1.1}
\begin{tabular}{lrrrrrrr}
\toprule
Portfolio & Return & Volatility & Sharpe & MDD & Calmar & Return/ave.DD & Turnover \\
\midrule\multicolumn{8}{l}{\textbf{In-sample period: gross of transaction costs}}\\
CNN-LSTM \ee\ & 0.3271 & 0.2753 & \textbf{1.1882} & 0.4089 & \textbf{0.7999} & 3.4053 & 0.4811 \\
CNN-LSTM TS & 0.2808 & 0.2843 & 0.9878 & 0.3849 & 0.7296 & 2.3143 & 0.8850 \\
RLinear \ee\ & 0.3144 & 0.3153 & 0.9969 & 0.4432 & 0.7093 & 3.3641 & 1.5285 \\
RLinear TS & 0.2901 & 0.3113 & 0.9319 & 0.3653 & 0.7939 & 3.7114 & 1.7788 \\
EW & 0.1161 & 0.1294 & 0.8969 & 0.1891 & 0.6138 & \textbf{5.1315} & 0.0213 \\
MV & 0.2400 & 0.2770 & 0.8663 & 0.3064 & 0.7833 & 2.8717 & 0.6425 \\

\midrule\multicolumn{8}{l}{\textbf{In-sample period: net of transaction costs}}\\
CNN-LSTM \ee\ & 0.2768 & 0.2757 & \textbf{1.0039} & 0.4479 & \textbf{0.6180} & 2.2633 & 0.4811 \\
CNN-LSTM TS & 0.1883 & 0.2851 & 0.6605 & 0.4575 & 0.4116 & 1.1009 & 0.8850 \\
RLinear \ee\ & 0.1552 & 0.3162 & 0.4908 & 0.5298 & 0.2929 & 1.2041 & 1.5285 \\
RLinear TS & 0.1041 & 0.3123 & 0.3333 & 0.4353 & 0.2391 & 0.8411 & 1.7788 \\
EW & 0.1138 & 0.1294 & 0.8797 & 0.1897 & 0.6001 & \textbf{4.9699} & 0.0213 \\
MV & 0.1728 & 0.2779 & 0.6218 & 0.3622 & 0.4771 & 1.5181 & 0.6425 \\

\midrule\multicolumn{8}{l}{\textbf{Out-of-sample period: gross of transaction costs}}\\
CNN-LSTM \ee\ & 0.6856 & 0.4782 & \textbf{1.4337} & 0.4202 & \textbf{1.6317} & \textbf{7.1594} & 0.4040 \\
CNN-LSTM TS & 0.6373 & 0.4806 & 1.3259 & 0.5116 & 1.2456 & 4.3035 & 0.7815 \\
RLinear \ee\ & 0.6286 & 0.4690 & 1.3404 & 0.4440 & 1.4158 & 4.5100 & 1.4454 \\
RLinear TS & 0.4779 & 0.4607 & 1.0375 & 0.4863 & 0.9828 & 2.8373 & 1.8711 \\
EW & 0.1641 & 0.1962 & 0.8367 & 0.3331 & 0.4927 & 3.2076 & 0.0271 \\
MV & 0.5371 & 0.4745 & 1.1319 & 0.4753 & 1.1299 & 3.3383 & 0.5616 \\

\midrule\multicolumn{8}{l}{\textbf{Out-of-sample period: net of transaction costs}}\\
CNN-LSTM \ee\ & 0.6434 & 0.4787 & \textbf{1.3441} & 0.4273 & \textbf{1.5059} & \textbf{6.3480} & 0.4040 \\
CNN-LSTM TS & 0.5556 & 0.4817 & 1.1534 & 0.5559 & 0.9994 & 3.1887 & 0.7815 \\
RLinear \ee\ & 0.4776 & 0.4703 & 1.0155 & 0.4805 & 0.9940 & 2.5303 & 1.4454 \\
RLinear TS & 0.2824 & 0.4621 & 0.6113 & 0.5978 & 0.4724 & 1.2102 & 1.8711 \\
EW & 0.1613 & 0.1961 & 0.8223 & 0.3334 & 0.4838 & 3.0917 & 0.0271 \\
MV & 0.4784 & 0.4754 & 1.0063 & 0.5087 & 0.9405 & 2.5809 & 0.5616 \\
\bottomrule
\end{tabular}
\end{table}

\Cref{fig:crsp-weekly-nav,tab:crsp-weekly-results} show that the advantage of integrated training persists in the larger universe and at the lower trading frequency. In sample and net of transaction costs, \ee\ increases the Sharpe ratio from \(0.6605\) to \(1.0039\) for CNN-LSTM and from \(0.3333\) to \(0.4908\) for RLinear, while reducing weekly turnover under both predictors.
The out-of-sample results are stronger. For CNN-LSTM, \ee\ increases the annualized net return from \(0.5556\) to \(0.6434\) and the Sharpe ratio from \(1.1534\) to \(1.3441\), while reducing maximum drawdown from \(0.5559\) to \(0.4273\) and turnover from \(0.7815\) to \(0.4040\). For RLinear, the corresponding net return rises from \(0.2824\) to \(0.4776\) and the Sharpe ratio from \(0.6113\) to \(1.0155\), with maximum drawdown declining from \(0.5978\) to \(0.4805\) and turnover from \(1.8711\) to \(1.4454\). RLinear \ee\ also remains competitive with MV, delivering nearly the same net return (\(0.4776\) versus \(0.4784\)) with a slightly higher Sharpe ratio and a lower maximum drawdown. Overall, the results show that the gains from integrated training are not confined to the original ETF universe, daily rebalancing, or a short planning horizon.

\subsection{Comparison of scalability}
To evaluate computational efficiency, we compare the proposed MDFP layer with two state-of-the-art differentiable convex optimizers: BPQP and CvxpyLayer. 
All models are trained end-to-end under the same experimental pipeline using the RLinear predictor. We report the average time per training epoch. We also ran CNN-LSTM and observed qualitatively similar scaling, but omit it here for brevity.
The problem size is controlled by the prediction horizon~$H$: increasing~$H$ expands the optimization dimension, as every additional 10-step horizon introduces 70 new decision variables into the underlying~$\arg\min$~problem. 
For gradient computation, the MDFP layer adopts a Neumann-series approximation of the implicit Jacobian inverse, with the residual norm capped at~$10^{-6}$ to ensure stable and accurate differentiation while maintaining low computational overhead.

\begin{figure}[h!]
  \centering
  \includegraphics[width=0.8\linewidth]{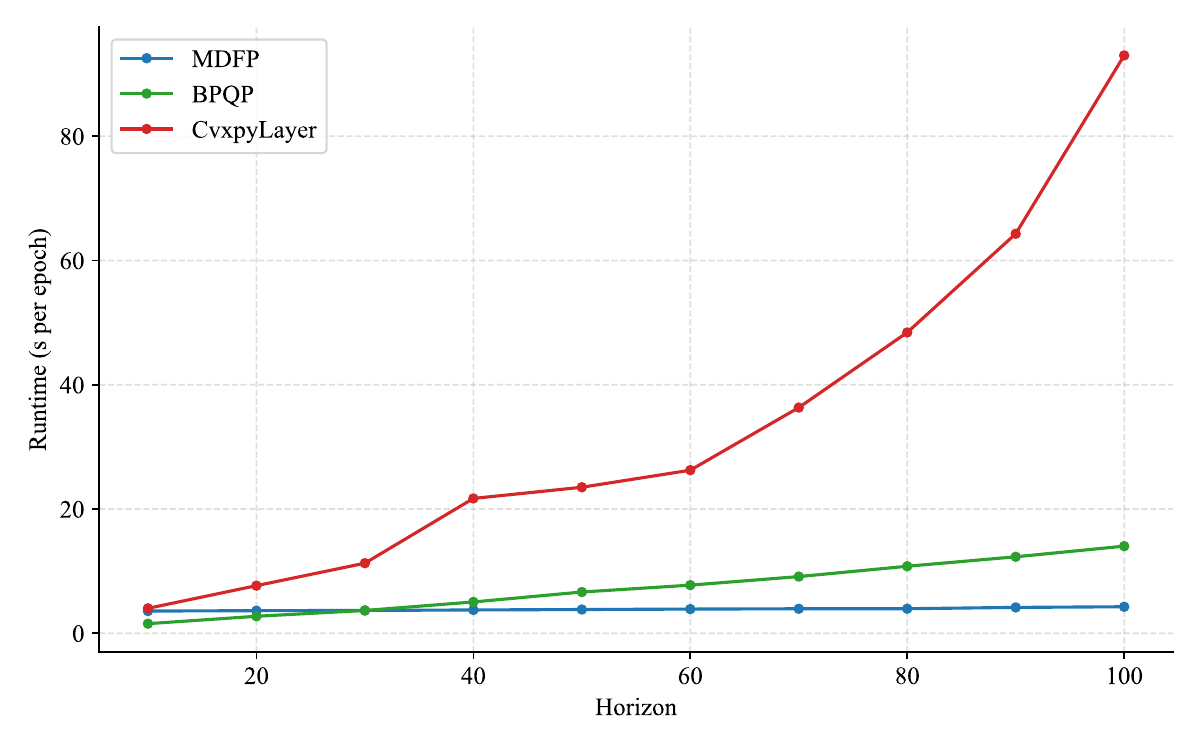}
  \caption{Runtime comparison across horizons based on 200 epochs, with lower values indicating better performance.}
  \label{fig:runtime}
\end{figure}

\revise{\Cref{fig:runtime} presents the runtime results for the seven-ETF setting. Over the tested horizon range, MDFP's average runtime per epoch increases by approximately \(0.69\,\mathrm{s}\), a smaller absolute increase than those of BPQP and CvxpyLayer.}
\revise{BPQP's runtime grows approximately linearly with the horizon in this experiment, while CvxpyLayer exhibits the steepest growth. At the largest tested horizon, MDFP is approximately \(3.29\times\) faster than BPQP and \(21.83\times\) faster than CvxpyLayer.}
These findings further corroborate our theoretical rationale. CvxpyLayer relies on large KKT factorizations and conic reformulations, while BPQP still incurs horizon-coupled linear-system solves whose iteration burden grows with size. \revise{MDFP uses a Neumann-series approximation that avoids explicit matrix inversion and large matrix factorizations, reducing runtime in the tested settings.}

\begin{figure}[h!]
  \centering
  \includegraphics[width=0.9\linewidth]{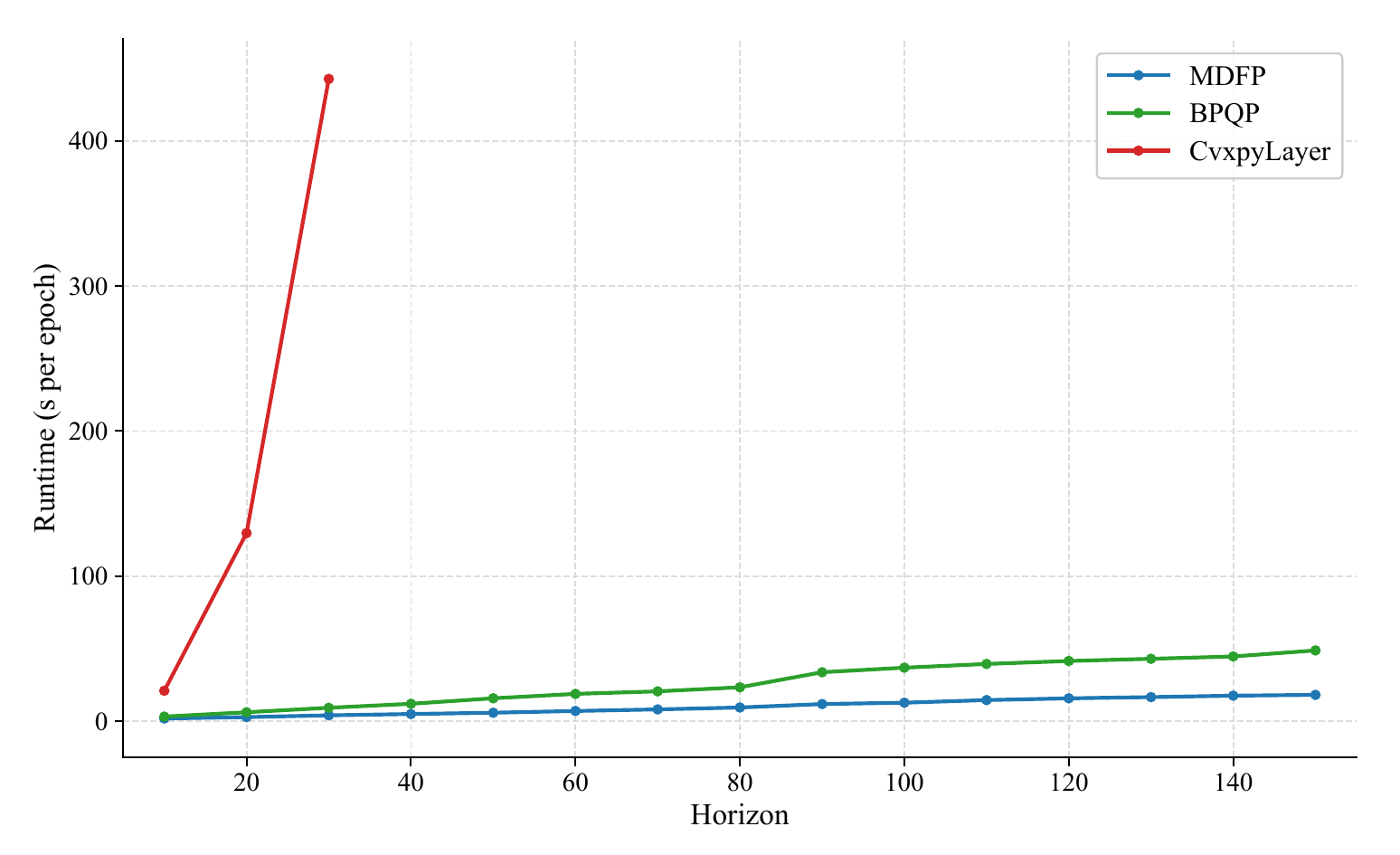}
  \caption{Layer-level runtime stress test with 100 assets across horizons based on 200 epochs, with lower values indicating better performance.}
  \label{fig:runtime-stress-100}
\end{figure}

We further examine whether the computational advantage persists when both the cross-sectional dimension and the planning horizon increase. The stress test uses \(N=100\) assets and extends the horizon to \(H=150\), with all differentiation methods evaluated under the same numerical tolerance of \(10^{-6}\). As shown in \cref{fig:runtime-stress-100}, MDFP runtime increases from \(1.91\,\mathrm{s}\) at \(H=10\) to \(18.19\,\mathrm{s}\) at \(H=150\), whereas BPQP increases from \(3.22\,\mathrm{s}\) to \(48.82\,\mathrm{s}\). At \(H=150\), BPQP therefore requires \(2.68\times\) the runtime of MDFP, corresponding to a \(62.7\%\) runtime reduction for MDFP. \revise{The approximately \(9.5\)-fold increase in MDFP runtime shows that its computational cost grows with the horizon. These comparisons support a lower measured runtime over the tested range.}

CvxpyLayer increases from \(21.04\,\mathrm{s}\) at \(H=10\) to \(442.71\,\mathrm{s}\) at \(H=30\), and no completed measurements were obtained beyond \(H=30\) in this setting. At \(H=30\), its runtime is already \(108.4\times\) that of MDFP and \(47.6\times\) that of BPQP. The stress test therefore shows that MDFP's scaling advantage is not confined to the original seven-ETF problem and remains pronounced when the asset universe and the horizon are enlarged simultaneously.

\section{Conclusion}

\label{section:conclusion}

In this work, we propose \ee\ for multi-period portfolio optimization, integrating prediction and decision-making within a single differentiable model. \revise{The predictor is trained using a mean-variance loss through a turnover-regularized allocation layer; performance net of transaction costs is assessed in the backtests.}
\revise{Over the full in-sample and out-of-sample periods in both the seven-ETF and dynamic 100-stock settings, \ee\ achieves higher net returns and Sharpe ratios and lower turnover than the corresponding two-stage models under both predictors.} With the RLinear model, \ee\ regularizes portfolio dynamics, yielding steadier adjustments and lower transaction intensity.
\revise{With the CNN-LSTM model, allocation paths exhibit extended holding periods interrupted by changes in portfolio composition. These full-period gains do not imply uniformly lower risk or superior performance in every market regime. In the seven-ETF experiment, CNN-LSTM \ee\ has higher volatility and maximum drawdown than TS over the full test period; in 2022, TS also achieves a higher net return and Sharpe ratio with lower volatility and maximum drawdown.}
Algorithmically, the proposed MDFP layer enables efficient and scalable training without explicit KKT factorizations and exhibits substantially better scaling with the horizon than the benchmark differentiable optimizers. This makes \ee\ readily applicable to realistic multi-period decision environments.

However, there are still some limitations to this study. The current MDFP formulation is developed under simplex constraints, where the mirror-descent geometry ensures a well-defined fixed point. Nevertheless, the underlying idea is more general: any optimization problem whose solution satisfies a stable fixed-point relation can, in principle, be embedded within the same \ee\ differentiation paradigm. Future work may formalize this generalization and extend it to richer constraint structures. Although the dynamic 100-stock experiment substantially broadens the empirical scope, further evaluation on still larger universes, additional asset classes, and alternative rebalancing frequencies would provide a more comprehensive assessment of robustness in high-dimensional decision environments. Furthermore, the CNN-LSTM does not provide a direct economic rationale for individual forecasts, so the allocation plots offer outcome-level rather than model-level transparency \citep{rudin2019stop}. Moreover, the experiments rely solely on historical returns as input features. Future work could incorporate macroeconomic, fundamental, and asset-specific information and investigate more transparent forecasting architectures \citep{gu2020empirical}. Advancing these directions would further consolidate \ee\ optimization as a general methodology for multi-period financial decision-making.

\section*{Acknowledgments}

This work was supported in part by the National Natural Science Foundation of China [Grants 12571325 and 72394364/72394360] and the Natural Science Foundation of Shanghai [Grant 24ZR1421300].

\appendix
\section{Proof of~\Cref{thm:md-fixedpoint}}\label{app:proof-fixedpoint}

\begin{proof}
Recall that $\Zbf_t^*=(z_{t+1}^*,\ldots,z_{t+H}^*)$ solves
$\min_{\Zbf_t\in\boldsymbol{\Omega}} f(\Zbf_t,\theta)$.
The KKT conditions yield, for each stage $s=t+1,\ldots,t+H$, the existence of multipliers 
$\mu_s\in\mathbb{R}$ and $\nu_s\in\mathbb{R}^N_{\ge0}$ such that
\begin{align*}
\nabla_{z_s} f(\Zbf_t^*, \theta) + \mu_s \mathbf{1} - \nu_s &= 0,\\
\nu_{s,i} &\ge 0,\quad i=1,\ldots,N,\\
\nu_{s,i} z^*_{s,i} &= 0,\quad i=1,\ldots,N,\\
\mathbf{1}^\top z_s^* &= 1,\quad z_s^*\ge 0.
\end{align*}
Let $\mathcal{S}_s=\{i:z^*_{s,i}>0\}$ be the support set at stage $s$.
By complementary slackness, $\nu_{s,i}=0$ and
$\nabla_{z_{s,i}} f(\Zbf_t^*,\theta)=-\mu_s$ for
$i\in\mathcal{S}_s$, while $z^*_{s,i}=0$ and $\nu_{s,i}\ge0$ for
$i\notin\mathcal{S}_s$.

Consider the stagewise Mirror Descent (MD) map:
\[
\Phi_{s,i}(\Zbf_t, \theta)
=\frac{z_{s,i}\exp\!\big(-\eta\nabla_{z_{s,i}} f(\Zbf_t, \theta)\big)}
{\sum_{j=1}^N z_{s,j}\exp\!\big(-\eta\nabla_{z_{s,j}} f(\Zbf_t, \theta)\big)},
\qquad \eta>0,
\]
and $\Phi(\Zbf_t, \theta)=(\Phi_{t+1}(\Zbf_t, \theta),\ldots,\Phi_{t+H}(\Zbf_t, \theta))$.
Evaluating at $\Zbf_t^*$ and distinguishing between
$i\in\mathcal{S}_s$ and $i\notin\mathcal{S}_s$:
\[
\text{numerator}=
\begin{cases}
z^*_{s,i}\exp(\eta\mu_s), & i\in\mathcal{S}_s,\\
0, & i\notin\mathcal{S}_s
\end{cases},
\]

\begin{align*}
\text{denominator}
&=\sum_{j\in\mathcal{S}_s} z^*_{s,j}\exp(\eta\mu_s)\\
&=\exp(\eta\mu_s)\sum_{j\in\mathcal{S}_s} z^*_{s,j}
=\exp(\eta\mu_s).
\end{align*}
Thus, $\Phi_{s,i}(\Zbf_t^*,\theta)=z_{s,i}^*$ for every $i$.
It follows that $\Phi_s(\Zbf_t^*,\theta)=z_s^*$ for each $s$, and hence
$\Phi(\Zbf_t^*,\theta)=\Zbf_t^*$.
\end{proof}

\section{Proof of~\Cref{prop:md-gradient}}
\label{app:proof-md-gradient}

\begin{proof}
Starting from the fixed-point relation 
$\Zbf_t^{*}=\Phi(\Zbf_t^{*},\theta)$,
take the total derivative with respect to~$\theta$:
\[
\frac{\partial \Zbf_t^{*}}{\partial \theta}
= \partial_{\Zbf_t^*}\Phi(\Zbf_t^{*},\theta)
  \frac{\partial \Zbf_t^{*}}{\partial \theta}
 + \frac{\partial \Phi(\Zbf_t^{*},\theta)}{\partial \theta}.
\]
Rearranging and using the assumed invertibility of
$I-\partial_{\Zbf_t^*}\Phi(\Zbf_t^*,\theta)$ gives
\[
\frac{\partial \Zbf_t^{*}}{\partial \theta}
=\big(I - \partial_{\Zbf_t^*}\Phi(\Zbf_t^{*},\theta)\big)^{-1}
\frac{\partial \Phi(\Zbf_t^{*},\theta)}{\partial \theta},
\]
which is exactly~\eqref{eq:md-gradient}.
\end{proof}

\section{Proof of~\Cref{thm:fp-kkt}}
\label{app:proof-fp-kkt}

\begin{proof}

Let $E=I_H\otimes\mathbf{1}_N\in\mathbb{R}^{HN\times H}$, so the equality constraints of the product simplex are $E^\top\Zbf_t=\mathbf{1}_H$.
Set $q(\Zbf_t,\theta)=\nabla_{\Zbf_t}\hat F_t(\Zbf_t,\theta)$,
$G=\nabla^2_{\Zbf_t\Zbf_t}\hat F_t(\Zbf_t^*,\theta)$, and
$K=\partial_\theta q(\Zbf_t^*,\theta)$.
Strong convexity of the inner objective makes $G$ symmetric positive definite.

We first consider an interior solution. Stagewise KKT stationarity yields scalars $c_s$ such that
\[
q_s(\Zbf_t^*,\theta)=c_s\mathbf{1}_N,
\qquad s=t+1,\ldots,t+H,
\]
where $q_s$ is the $s$-th stage block of $q$.
Define $a_s(\Zbf_t,\theta):=z_s\odot\exp(-\eta q_s(\Zbf_t,\theta))$
and $d_s(\Zbf_t,\theta):=\mathbf{1}_N^\top a_s(\Zbf_t,\theta)$, so that
$\Phi_s=a_s/d_s$. At the optimum,
$a_s(\Zbf_t^*,\theta)=e^{-\eta c_s}z_s^*$ and
$d_s(\Zbf_t^*,\theta)=e^{-\eta c_s}$.

Write $G_{sr}=\nabla^2_{z_s z_r}\hat F_t(\Zbf_t^*,\theta)$,
$P_s=I_N-z_s^*\mathbf{1}_N^\top$, and
$W_s=\operatorname{diag}(z_s^*)$.

  Each stagewise mean--variance term involves only the variable $z_s$, while each turnover term links only neighboring allocations; thus, the mixed partial $G_{sr}$ vanishes whenever $|s - r| > 1$, which implies that the Hessian $G$ assumes a block tridiagonal structure.

Writing $q_s^*=q_s(\Zbf_t^*,\theta)$, the product and chain rules yield
\[
\begin{aligned}
\left.\frac{\partial a_s}{\partial z_r}\right|_{\Zbf_t^*}
&=
\operatorname{diag}\bigl(\exp(-\eta q_s^*)\bigr)
\frac{\partial z_s}{\partial z_r}
+
\operatorname{diag}(z_s^*)
\left.\frac{\partial\exp(-\eta q_s)}{\partial z_r}
\right|_{\Zbf_t^*}\\
&=
\delta_{sr}\operatorname{diag}\bigl(\exp(-\eta q_s^*)\bigr)
-
\eta\operatorname{diag}(z_s^*)
\operatorname{diag}\bigl(\exp(-\eta q_s^*)\bigr)G_{sr}\\
&=
\delta_{sr}\operatorname{diag}\bigl(\exp(-\eta q_s^*)\bigr)
-
\eta\operatorname{diag}\bigl(z_s^*\odot\exp(-\eta q_s^*)\bigr)G_{sr}\\
&=
e^{-\eta c_s}
\bigl(\delta_{sr}I_N-\eta W_sG_{sr}\bigr),
\end{aligned}
\]
where $\delta_{sr}$ is the Kronecker delta.
Since $d_s=\mathbf{1}_N^\top a_s$, the quotient rule gives
\[
\frac{\partial\Phi_s}{\partial z_r}
=
\frac{1}{d_s}
\bigl(I_N-\Phi_s\mathbf{1}_N^\top\bigr)
\frac{\partial a_s}{\partial z_r}.
\]
Evaluating at $\Zbf_t^*$ yields
\begin{equation}
\label{eq:global-Phi-blocks}
\left.\frac{\partial\Phi_s}{\partial z_r}\right|_{\Zbf_t^*}
=
\begin{cases}
P_s\bigl(I_N-\eta W_sG_{ss}\bigr), & r=s,\\[2pt]
-\eta P_sW_sG_{sr}, & r\ne s.
\end{cases}
\end{equation}
It follows that $\partial\Phi_s/\partial z_r=0$ whenever $|s-r|>1$, so the fixed-point Jacobian is also block tridiagonal.

We now collect the stagewise derivatives and compare the fixed-point sensitivity with the KKT sensitivity.
Let $W=\operatorname{diag}(\Zbf_t^*)$ and
$P=I_{HN}-WEE^\top=\operatorname{blkdiag}(P_{t+1},\ldots,P_{t+H})$.
Because every stage allocation sums to one, $E^\top W E=I_H$ and $E^\top P=0$.
Thus, $P$ projects onto the tangent space
$\mathcal{T}=\{v\in\mathbb{R}^{HN}:E^\top v=0\}$ and satisfies $Pv=v$ for every $v\in\mathcal{T}$.
Writing $K_s$ for the $s$-th stage block of $K$, the quotient rule gives
\[
\begin{aligned}
\left.\frac{\partial\Phi_s}{\partial\theta}\right|_{\Zbf_t^*}
&=\frac{1}{d_s^*}P_s
\left.\frac{\partial a_s}{\partial\theta}\right|_{\Zbf_t^*}\\
&=\frac{1}{e^{-\eta c_s}}P_s
\left(-\eta e^{-\eta c_s}W_sK_s\right)\\
&=-\eta P_sW_sK_s.
\end{aligned}
\]
Collecting this result and the blocks in~\eqref{eq:global-Phi-blocks} gives
\begin{equation}
\label{eq:global-fixed-point-Jacobian}
\begin{aligned}
J:=\partial_{\Zbf_t}\Phi(\Zbf_t^*,\theta)
&=P(I_{HN}-\eta WG),\\
\partial_\theta\Phi(\Zbf_t^*,\theta)&=-\eta PWK.
\end{aligned}
\end{equation}

Differentiating $\Zbf_t^*=\Phi(\Zbf_t^*,\theta)$ and using~\eqref{eq:global-fixed-point-Jacobian} gives
\begin{equation}
\label{eq:global-fixed-point-sensitivity}
\bigl(I_{HN}-P+\eta PWG\bigr)
\frac{\partial\Zbf_t^*}{\partial\theta}
=-\eta PWK.
\end{equation}
Because $\Phi(\Zbf_t,\theta)$ lies in the product simplex for every $(\Zbf_t,\theta)$,
differentiating $E^\top\Zbf_t^*=\mathbf{1}_H$ gives
$E^\top\frac{\partial\Zbf_t^*}{\partial\theta}=0$.
Consequently,
$P\frac{\partial\Zbf_t^*}{\partial\theta}
=
\frac{\partial\Zbf_t^*}{\partial\theta}$,
so~\eqref{eq:global-fixed-point-sensitivity} reduces to
$PW\left(G\frac{\partial\Zbf_t^*}{\partial\theta}+K\right)=0$.
Since the solution is interior, $W$ is invertible and
$\ker(PW)=\operatorname{range}(E)$ because
$\ker(P)=\operatorname{range}(WE)$.
There is therefore a matrix $M$ such that
$G\frac{\partial\Zbf_t^*}{\partial\theta}+K=EM$ and
$E^\top\frac{\partial\Zbf_t^*}{\partial\theta}=0$.
The KKT conditions for the interior problem are
$q(\Zbf_t^*,\theta)+E\mu=0$ and $E^\top\Zbf_t^*=\mathbf{1}_H$.
Their derivatives are
$G\frac{\partial\Zbf_t^*}{\partial\theta}
+K+E\,\frac{\partial\mu}{\partial\theta}=0$ and
$E^\top\frac{\partial\Zbf_t^*}{\partial\theta}=0$,
which are precisely the preceding relations after setting
$M=-\partial\mu/\partial\theta$.
Thus, the fixed-point and KKT sensitivities coincide for the complete coupled problem.

It remains to consider a boundary solution.
For each stage, let
$\mathcal{S}_s=\{i:z_{s,i}^*>0\}$ and
$\mathcal{A}_s=\{i:z_{s,i}^*=0\}$.
Define the horizon-wide coordinate sets
$\mathcal{S}=\{\,(s,i):s=t+1,\ldots,t+H,\ i\in\mathcal{S}_s\,\}$ and
$\mathcal{A}=\{\,(s,i):s=t+1,\ldots,t+H,\ i\in\mathcal{A}_s\,\}$.
Strict complementarity implies that these sets are locally constant.
Let $c_s$ be the common value of $q_{s,i}(\Zbf_t^*,\theta)$ on
$\mathcal{S}_s$.  For every $i\in\mathcal{A}_s$,
strict complementarity gives $q_{s,i}(\Zbf_t^*,\theta)-c_s>0$.
Because the numerator of $\Phi_{s,i}$ contains the multiplicative factor
$z_{s,i}$, its derivatives at the fixed point satisfy
$\partial\Phi_{s,i}/\partial z_{r,j}=0$ whenever
$(r,j)\ne(s,i)$ and $\partial_\theta\Phi_{s,i}=0$, whereas its self-derivative is
$\exp(-\eta[q_{s,i}(\Zbf_t^*,\theta)-c_s])\in(0,1)$.
The differentiated fixed-point equation therefore implies
$\frac{\partial(\Zbf_t^*)_{\mathcal{A}}}{\partial\theta}=0$.

After restricting to the free coordinates, the mapping is the same normalized multiplicative map on a product of lower-dimensional simplices.
Let $W_{\mathcal{S}}$, $E_{\mathcal{S}}$,
$G_{\mathcal{S}\mathcal{S}}$, and $K_{\mathcal{S}}$
denote the corresponding restrictions.
The principal submatrix $G_{\mathcal{S}\mathcal{S}}$ is positive definite.
Applying the interior argument to the reduced system gives
$G_{\mathcal{S}\mathcal{S}}
\frac{\partial(\Zbf_t^*)_{\mathcal{S}}}{\partial\theta}
+K_{\mathcal{S}}
=E_{\mathcal{S}}M$,
$E_{\mathcal{S}}^\top
\frac{\partial(\Zbf_t^*)_{\mathcal{S}}}{\partial\theta}
=0$, and
$\frac{\partial(\Zbf_t^*)_{\mathcal{A}}}{\partial\theta}=0$.
These are precisely the differentiated KKT conditions on the locally constant active set.
Therefore, under strict complementarity, the fixed-point sensitivity is identical to the KKT sensitivity for both interior and boundary solutions of the full multi-stage problem.
\end{proof}

\section{Proof of~\Cref{prop:mdfp-spectral-radius}}\label{app:proof-spectral-radius}

\begin{proof}
We use the global notation introduced in the proof of~\cref{thm:fp-kkt}.
First suppose that $\Zbf_t^*$ is an interior solution.
Recall that $J=P(I_{HN}-\eta WG)$, $P=I_{HN}-WEE^\top$, and
$\mathcal{T}=\ker(E^\top)$.
Because $E^\top P=0$, the range of $J$ is contained in $\mathcal{T}$.
Consequently, if $Jv=\lambda v$ with $\lambda\ne0$, then $v\in\mathcal{T}$.
All remaining eigenvalues associated with directions outside the tangent space are zero.

For $v\in\mathcal{T}$, define $L=PWG$; then $Pv=v$ and
$Jv=v-\eta Lv$.
Equip $\mathcal{T}$ with the weighted inner product
$\langle u,v\rangle_{W^{-1}}=u^\top W^{-1}v$.
For any $u,v\in\mathcal{T}$,
\begin{align}
\langle u,Lv\rangle_{W^{-1}}
&=
u^\top W^{-1}PWGv \notag\\
&=
u^\top\bigl(I_{HN}-EE^\top W\bigr)Gv \notag\\
&=
u^\top Gv,
\label{eq:weighted-L}
\end{align}
where the last equality follows from $u^\top E=0$.
Since $G$ is symmetric,
$\langle u,Lv\rangle_{W^{-1}}=\langle Lu,v\rangle_{W^{-1}}$;
moreover, $\langle v,Lv\rangle_{W^{-1}}=v^\top Gv>0$ for every nonzero
$v\in\mathcal{T}$.
Thus, $L$ is self-adjoint and positive definite on $\mathcal{T}$ under the
$W^{-1}$-weighted inner product.  Its eigenvalues
$\tau_1,\ldots,\tau_{HN-H}$ are therefore real and strictly positive.
The eigenvalues of $J$ associated with tangent-space directions are $1-\eta\tau_i$,
$i=1,\ldots,HN-H$. It follows that
\begin{equation}
\label{eq:global-eta-bound}
0<\eta<\frac{2}{\tau_{\max}},
\qquad
\tau_{\max}:=\max_i\tau_i,
\end{equation}
implies $\rho(J)<1$.
A convenient sufficient bound follows from the Rayleigh quotient.
Since
$\langle v,Lv\rangle_{W^{-1}}=v^\top Gv$
and
$\langle v,v\rangle_{W^{-1}}=v^\top W^{-1}v$,
and since
$v^\top Gv\le\norm{G}_2\norm{v}_2^2$
and
$v^\top W^{-1}v\ge\norm{v}_2^2/\norm{W}_2$
for every $v\neq0$,
\[
\begin{aligned}
\tau_{\max}
&=
\max
\frac{v^\top Gv}{v^\top W^{-1}v}
\le
\frac{\norm{G}_2\norm{v}_2^2}{\norm{v}_2^2/\norm{W}_2}
=
\norm{G}_2\norm{W}_2.
\end{aligned}
\]
Consequently, the explicit step-size restriction
$0<\eta<2/(\norm{G}_2\norm{W}_2)$
is sufficient to ensure $\rho(J)<1$.

We next consider a boundary solution under strict complementarity.
Using the free and active index sets
$\mathcal{S}$ and $\mathcal{A}$ defined in the proof of~\cref{thm:fp-kkt},
and ordering the free coordinates first, the Jacobian has the block upper-triangular form
\[
J=
\begin{pmatrix}
J_{\mathcal{S}\mathcal{S}} &
J_{\mathcal{S}\mathcal{A}}\\
0 &
J_{\mathcal{A}\mathcal{A}}
\end{pmatrix}.
\]
The free-coordinate block $J_{\mathcal{S}\mathcal{S}}$ is the Jacobian of the normalized multiplicative map on the reduced product simplex.
The reduced Hessian $G_{\mathcal{S}\mathcal{S}}$ is positive definite, so the preceding weighted-inner-product argument applies with
$W_{\mathcal{S}}$, $E_{\mathcal{S}}$, and
$G_{\mathcal{S}\mathcal{S}}$.
Accordingly, $\rho(J_{\mathcal{S}\mathcal{S}})<1$ for a sufficiently small positive step size.

For an active coordinate $i\in\mathcal{A}_s$, write
$\Phi_{s,i}=n_{s,i}/d_s$, where
$n_{s,i}=z_{s,i}\exp(-\eta q_{s,i})$.
At the fixed point, $n_{s,i}^*=0$ and
$d_s^*=\sum_{j\in\mathcal{S}_s}z_{s,j}^*e^{-\eta c_s}=e^{-\eta c_s}$.
Let $q_{s,i}^*:=q_{s,i}(\Zbf_t^*,\theta)$.
Since $z_{s,i}^*=0$, differentiating the numerator gives
\[
\begin{aligned}
\left.\frac{\partial n_{s,i}}{\partial z_{r,j}}\right|_{\Zbf_t^*}
&=
\delta_{sr}\delta_{ij}e^{-\eta q_{s,i}^*}
-\eta z_{s,i}^*e^{-\eta q_{s,i}^*}
\left.\frac{\partial q_{s,i}}{\partial z_{r,j}}\right|_{\Zbf_t^*}
=
\delta_{sr}\delta_{ij}e^{-\eta q_{s,i}^*}.
\end{aligned}
\]
The quotient rule therefore gives
\[
\left.\frac{\partial\Phi_{s,i}}{\partial z_{r,j}}\right|_{\Zbf_t^*}
=
\begin{cases}
\exp\bigl(-\eta(q_{s,i}^*-c_s)\bigr), & (r,j)=(s,i),\\
0, & (r,j)\ne(s,i).
\end{cases}
\]
Strict complementarity implies $q_{s,i}^*-c_s>0$.
Hence, $J_{\mathcal{A}\mathcal{A}}$ is diagonal with entries in $(0,1)$, and
$\rho(J_{\mathcal{A}\mathcal{A}})<1$ for every $\eta>0$.
Since the eigenvalues of a block upper-triangular matrix are the union of the eigenvalues of its diagonal blocks,
$\rho(J)=\max\{\rho(J_{\mathcal{S}\mathcal{S}}),
\rho(J_{\mathcal{A}\mathcal{A}})\}<1$ for a sufficiently small positive step size.
This proves the proposition for both interior and boundary solutions.
\end{proof}

\bibliographystyle{plainnat}
\bibliography{references}

\end{document}